\documentclass[twocolumn,prd,nofootinbib,aps,floats,floatfix,amsmath,amssymb,secnumarabic,superscriptaddress]{revtex4-1} 

\usepackage[utf8]{inputenc}
\usepackage[margin=1in]{geometry}
\usepackage{amsmath,amssymb}
\usepackage{graphicx}
\usepackage{xcolor}
\usepackage[nice]{nicefrac}
\usepackage{natbib}
\usepackage{verbatim}
\usepackage{longtable}

\def\sfrac#1#2{{\textstyle{#1\over #2}}}
\newcommand{\be}{\begin{equation}}
\newcommand{\ee}{\end{equation}}
\newcommand{\ba}{\begin{array}}
\newcommand{\ea}{\end{array}}
\newcommand{\bea}{\begin{eqnarray}}
\newcommand{\eea}{\end{eqnarray}}
\newcommand{\sss}{\scriptscriptstyle}
\newcommand{\W}{{\sss W}}
\newcommand{\nn}{\nonumber}
\def\jc#1{{\color{red}JC: #1}}

\begin{document}

\title{Wall speed and shape in singlet-assisted\\ strong
electroweak phase transitions}
\author{Avi Friedlander}
\email{avi.friedlander@queensu.ca}
\affiliation{Queen's University, Department of Physics \& Engineering Physics Astronomy
Kingston, Ontario, K7L 3N6 Kingston, Canada}
\author{Ian Banta}
\email{banta@physics.ucsb.edu}
\affiliation{Department of Physics, Williams College, Williamstown, MA 01267}
\author{James M.\ Cline}
\email{jcline@physics.mcgill.ca}
\affiliation{McGill University, Department of Physics, 3600 University St., Montr\'eal, QC H3A2T8 Canada}
\author{David Tucker-Smith}\email{dtuckers@williams.edu}
\affiliation{Department of Physics, Williams College, Williamstown, MA 01267}

\date{August 2020}
\begin{abstract}
    Models with singlet fields coupling to the Higgs can enable a strongly first order electroweak phase transition, of interest for baryogenesis and gravity waves.  We improve on previous attempts to self-consistently solve for the bubble wall properties---wall speed $v_w$ and shape---in a highly predictive class of models with $Z_2$-symmetric singlet potentials.  A new algorithm is implemented to determine $v_w$ and the wall profiles throughout the singlet parameter space in the case of subsonic walls, focusing on models with strong enough phase transitions to satisfy the sphaleron washout constraint for electroweak baryogenesis.  
    We find speeds as low as $v_w \cong 0.1$ in our scan over parameter space,
     and the
singlet must be relatively light to have a subsonic wall, $m_s \lesssim 135$ GeV.
\end{abstract}
\maketitle

\section{Introduction}

The electroweak phase transition (EWPT) in the early universe has been intensively studied as a possible source for the cosmic baryon asymmetry and gravitational waves. Within the standard model (SM) neither of these interesting outcomes are possible, given the known mass of the Higgs boson, because the phase transition is a smooth crossover \cite{Kajantie:1996qd,Kajantie:1996mn}, whereas a first-order EWPT is required for electroweak baryogenesis (EWBG) and production of observable gravity waves (for reviews, see for example refs.\ \cite{Morrissey:2012db,Caprini:2019egz}).  New physics, typically in the form of scalar fields coupling to the Higgs boson, can however lead to a first-order transition, with consequent nucleation of bubbles of the true (electroweak symmetry broken) vacuum, at the onset of the transition.

In order to make quantitative predictions for either baryogenesis or gravitational wave production in a given model,
it is necessary to understand the detailed properties of the
phase transition bubbles, especially the shape of the bubble walls (typically modeled as a tanh with some thickness $L_w$) and the terminal velocity $v_w$ attained by them, once the forces of internal pressure and external friction from the plasma have balanced each other.
This calculation, first carried out for the SM in refs.\ \cite{Liu:1992tn,Moore:1995ua,Moore:1995si} (assuming a light Higgs boson), and in the Minimal Supersymmetric Standard Model, where the phase transition is enhanced by 
light stops, in refs.\ \cite{John:2000zq,Huber:2011aa}, turns out to be quite challenging because the frictional force, which requires solving the Boltzmann equations for the perturbations of the plasma caused by the wall, depends on the same wall properties that one is trying to determine.  

A self-consistent procedure to solve this system is numerically expensive, and for this reason many studies of EWBG or gravitational wave production leave $L_w$ and $v_w$ as phenomenological parameters that can be freely varied,
or in a somewhat better approximation, calculated by modeling the friction in a phenomenological way \cite{Espinosa:2010hh,Huber:2013kj,Megevand:2013hwa}.
However to assess the prospects for a specfic model
to yield interesting results, one must eventually carry out
the actual computation of $L_w$ and $v_w$ with the actual friction term derived from the fluid perturbations.  Accurate estimates of these parameters are needed to make quantitative predictions for baryogenesis or gravity wave production.

The  procedure becomes even more laborious in the case where an extra singlet field couples to the Higgs, in order to facilitate the first order transition, and also gets a vacuum expectation value (VEV) in the bubble wall
\cite{Profumo:2007wc,Espinosa:2011eu,Espinosa:2011ax}.  In that case one must solve for both field profiles, which has been attempted in refs.\ \cite{Konstandin:2014zta,Kozaczuk:2015owa}, subject to some limiting approximations.  In particular, these previous works assumed that the bubble wall shapes are described by tanh profiles.  In reality, the Higgs field and the singlet can have shapes that differ from such an assumption, and it is not obvious how strongly this affects the determination of $v_w$.  One of our main
purposes is to overcome this limitation by developing an algorithm to determine the actual wall profiles along with $v_w$.  In this work we restrict our investigation to subsonic wall speeds; for recent progress on highly relativistic walls, see 
Ref.\ \cite{Hoeche:2020rsg}.

Moreover previous studies of singlet-assisted strong EWPTs have focused on a few benchmark models. In the present work we make a comprehensive scan of the parameter space for a class of models, where the singlet potential has the $Z_2$ symmetry $s\to -s$,
and the singlet VEV disappears at low temperatures.
This choice has the virtue of simplicity, being characterized by three parameters, the singlet
mass $m_s$, its cross coupling $\lambda_{hs}$ to the Higgs, and the VEV $w_0$ of $s$ in the false vacuum where $h = 0$.  The barrier between the true and false vacua provided by the $\lambda_{hs}h^2 s^2$ interaction is already present at tree level, and
is what enables the phase transition to be strongly 
first order \cite{Espinosa:2011eu,Espinosa:2011ax}.  Moreover with $\langle s\rangle=0$ at $T=0$, the new sources of CP violation needed for EWBG are not overly 
constrained by experimental limits on electric dipole moments.

The paper is organized as follows.
Section \ref{sec:model} describes the singlet scalar model used throughout the paper. Section \ref{sec:transition} outlines the main features of the electroweak phase transition dynamics that will be studied in detail in the following. In section \ref{sec:dynamics} the methodology for determining the wall dynamics, including its velocity, are described; the results of those calculations are presented in section \ref{sec:results}. Conclusions are given in section \ref{sec:conclusion}.  Appendices contain details concerning the finite-temperature effective potential (appendix \ref{appA}) and diffusion equations used to determine the fluid perturbations
(appendix \ref{sec:AppendBoltz}).

\section{The Model}\label{sec:model}

A simple extension of the SM is the addition of a scalar
singlet $s$  that couples only to the Higgs field, and has
the $Z_2$ symmetry $s\to -s$. Its  zero-temperature tree level potential is given by
\bea \label{eq:treeLevel}
    V_0 &=& \lambda_h\left(|H|^2 - \frac{1}{2}v_0^2\right)^2 + \frac{1}{4}\lambda_s\left(s^2 - w_0^2\right)^2\nn\\
    &+& \frac{1}{2}\lambda_{hs}|H|^2s^2 
\eea
where $H$ is the Standard Model Higgs doublet, and  $\lambda_h$, $v_0$ are the Higgs self-coupling and VEV respectively. There are three new parameters
$\lambda_s$, $w_0$, and $\lambda_{hs}$, that describe the singlet's self coupling, its VEV when in the false minimum where  $H=0$, and the coupling between $H$ and $s$.  There is no loss of generality by omitting a separate $m_0^2 s^2$ mass term.  The physical singlet mass in the electroweak broken vacuum is given by 
\be
   m_s^2 = -\lambda_s w_0^2 + \sfrac12\lambda_{hs} v_0^2
   \label{smass}
\ee
We restrict the parameters so that $m_s^2>0$, implying that $\langle s\rangle = 0$ in the true vacuum.
The Higgs doublet components are 
\begin{equation}
    H = (\chi_1 + i \chi_2,\ h + i \chi_3)^T /\sqrt{2} 
\end{equation}
where $h$ denotes the background Higgs field, and the $\chi$'s are the Goldstone bosons.

The full effective potential takes into account one-loop corrections and temperature effects, 
\begin{equation}
    V_{\rm eff} = V_0 + V_1 + V_{CT} + V_T
\end{equation}
where $V_0$ is the tree-level potential (\ref{eq:treeLevel}), $V_1$ is the one-loop correction, $V_{CT}$ contains the counterterms associated with $V_1$, and $V_T$ is the thermal contribution, including ring resummation of thermal masses.  These expressions are standard, and we have 
described them in detail in appendix \ref{appA}.
$V_{\rm eff}$ is determined by the measured SM parameters  and the three new ones, that we henceforth take to be $w_0$, $\lambda_{hs}$  and the singlet mass $m_s$ by trading
$\lambda_s$ for $m_s$ through eq.\ (\ref{smass}).

\begin{figure}[b]
	\centering
	\includegraphics[width=0.375\textwidth]{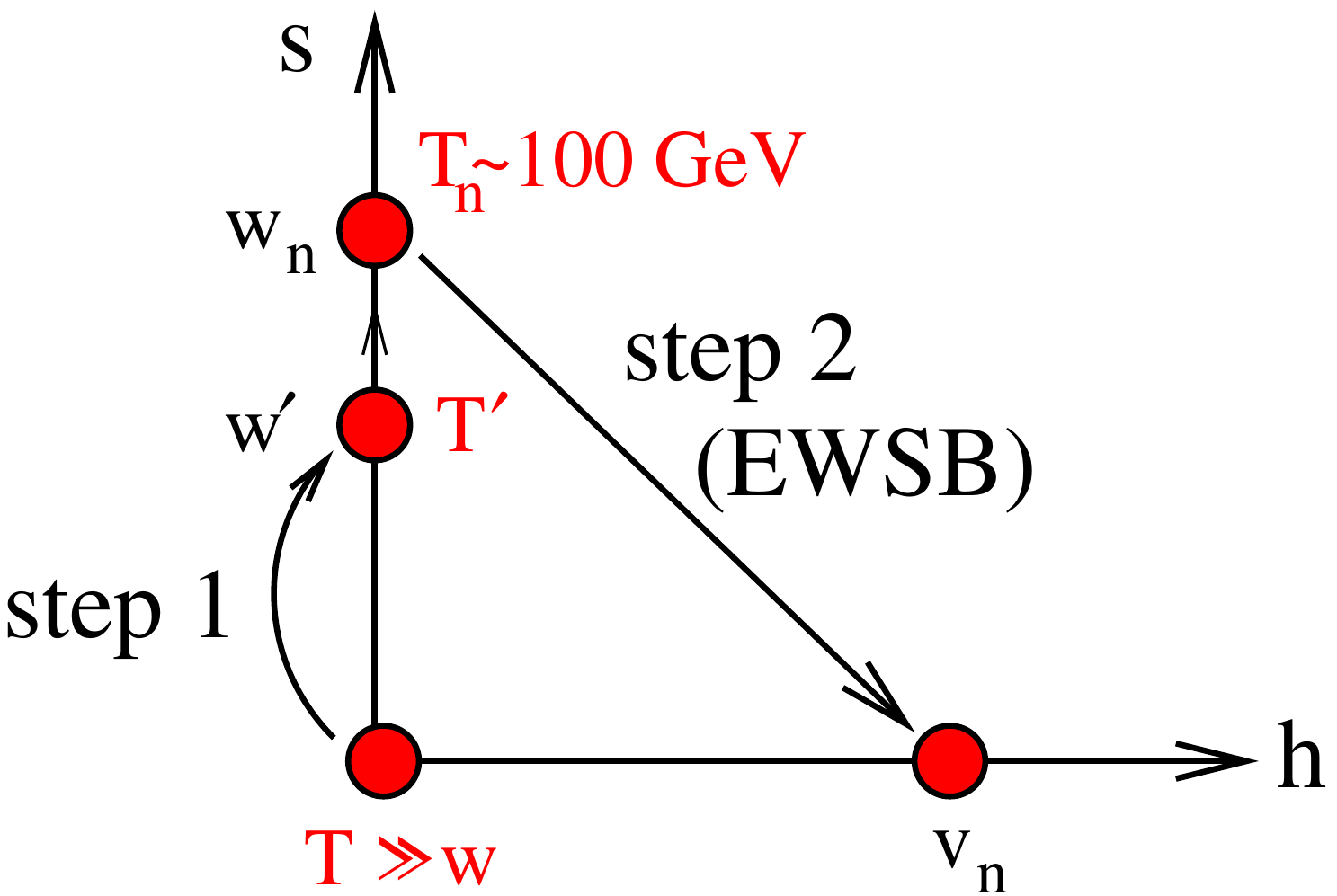}
	\caption{Sequence of phase transitions in field space.}
	\label{fig:2step}
\end{figure}

\section{Phase Transition} \label{sec:transition}

Because of our assumption that $\langle s\rangle =0$ at low temperatures, the $\lambda_{hs}h^2s^2$ coupling creates a barrier in field space between the false and true vacua, and gives rise to a two-step phase transition.
At temperatures $T\gg w$, the minimum of the potential is at the origin, where both electroweak symmetry and the $Z_2$ symmetry are restored. At some temperature $T'$, the first transition occurs, where $\langle s\rangle \to  w'$ (see fig.\ \ref{fig:2step}). As $T$ decreases, $\langle s\rangle$ and the corresponding minimum of the potential becomes metastable. At the critical temperature $T_c$, the two minima become degenerate,
and at a slightly lower temperature $T_n$ nucleation
of bubbles begins, signaling the 
second transition where electroweak symmetry is broken while the $Z_2$ symmetry is restored. These two transitions are summarized as
\begin{enumerate}
    \item At $T=T'$ $(h,s):(0,0) \rightarrow (0,w')$
    \item At $T=T_n$ $(h,s):(0,w_n) \rightarrow (v_n,0)$\,.
\end{enumerate}

It is the second transition that is important for baryogenesis. We note that while domain walls could form during the first transition, the restoration of the $Z_2$ symmetry during the second transition will cause them to annihilate. This occurs long before they can dominate the energy density of the universe, hence we expect that no cosmological problems will arise from this brief appearance of domain walls \cite{Espinosa:2011eu}.

\begin{figure*}[htb]
	\centering
	\includegraphics[width=0.8\textwidth]{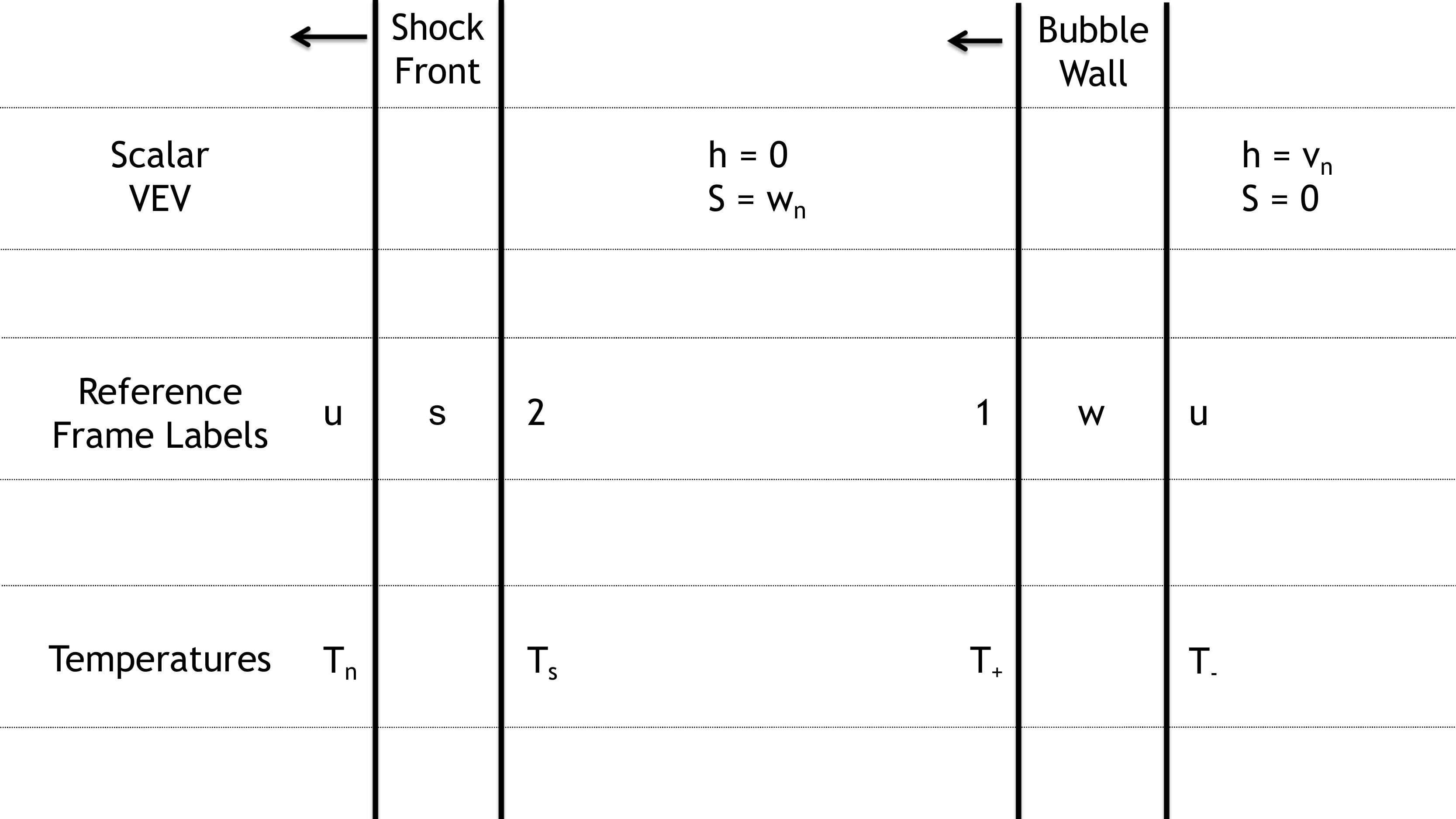}
	\caption{Illustration of the geometry of a deflagration. The bubble wall and shock front are moving to the left with the inside of the bubble being on the right of the figure.}
	\label{fig:DeflagrationLabels}
\end{figure*}

The dynamics of the phase transition depend strongly
on $T_n$, 
the temperature at which the probability of a bubble nucleating within one Hubble volume per Hubble time is $\mathcal{O}(1)$ \cite{Quiros:1999jp}. The tunneling probability goes as $\exp(-S_3/T)$, where $S_3$ is the three-dimensional Euclidean action. We used CosmoTransitions \cite{Wainwright:2011kj} 
to find phase transition candidates and to determine $T_n$\footnote{CosmoTransitions 
occasionally fails to find transitions when they should exist; in such cases,
changing the value of $\lambda_{hs}$ by 
$O(10^{-6})$ can overcome the problem.
Moreover, CosmoTransitions often reports more phase transitions than expected for a given model; we find that defining the EWPT
as  the most recent first order transition where the Higgs' VEV in the unbroken phase is smaller than the nucleation temperature gives correct results.    Of those, only phase transitions that ended with no singlet VEV were studied. We found it a useful cross-check to require the transition identified by CosmoTransitions to have a critical temperature that matched our own calculations for the parameter point in question.
Ref.\ \cite{Baum:2020vfl} has recently emphasized the importance of accurately determining the  nucleation temperature in order to reliably characterize the nature of the transition.}.  The criterion for the nucleation temperature is
taken to be $S_3/T_n=140$ \cite{Quiros:1999jp}.  

Since our investigation is motivated by electroweak baryogenesis,
we focus attention on first order transitions that are strong enough to preserve the baryon asymmetry from washout by residual sphaleron interaction inside the bubbles. This requires the Higgs VEV at the nucleation temperature to satisfy \cite{Moore:1998swa}
\be
    {v_n\over T_n} \gtrsim 1.1\, .
    \label{sph_bound}
\ee

\section{Wall Dynamics} \label{sec:dynamics}

The bubble-wall dynamics are determined by the interactions of the 
Higgs and singlet fields with a thermal fluid consisting of top quarks,
electroweak gauge bosons, and any other particles to which the scalars 
couple significantly.  After the bubble nucleates, it expands 
due to the outward pressure caused by the potential difference between the phases of the scalar fields on either side.  The interactions of the wall with the surrounding fluid counteract the expansion by a friction force that depends on the speed and shape of the wall. 

If the friction is strong  enough,  the  bubbles  reach  a  steady-state  velocity  whose  value  is  relevant  for  gravitational waves and baryogenesis.  The terminal $v_w$ depends upon the field profiles 
that solve the equations of motion.  These in turn depend  upon  the  temperature
$T_w$ of  the  wall and the $v_w$-dependent friction exerted by the plasma on the wall, leading to $T_w>T_n$, due to heating by the fluid.  A self-consistent solution thus requires simultaneously solving for $v_w$ and the 
scalar field profiles in the wall.

\subsection{Deflagration profiles} \label{sec:Temp}
The wall temperature, and the rest of the dynamics of the bubble, depend on 
whether the phase transition proceeds through deflagrations or detonations
(hybrids of these two are also possible).
For subsonic bubble walls, with $v_w < 1/\sqrt{3}$, the bubbles grow via deflagrations \cite{Espinosa:2010hh}, in which the wall is preceded by a shock front that moves through the fluid, perturbing it, increasing the temperature from $T_n$ to $T_s$ and causing
the wall to move (see fig.\ \ref{fig:DeflagrationLabels}). The fluid velocity decreases until the point where the wall passes it, so that the fluid behind the wall is at rest relative to that preceding the shock front.  In this
work we limit our investigation to
the case of deflagrations, hence subsonic walls, deferring the study of supersonic
walls to the future \cite{WIP}.

Because of the heating, the bubble wall dynamics are not determined at temperature $T_n$, but rather the temperature of the fluid near the wall. The calculation is performed at times sufficiently long after nucleation that the bubble has reached a steady-state velocity, and the profiles of the fluid perturbations vary on scales much larger than the wall thickness.  Therefore one approximates the wall as a discontinuity, such that the fluid temperature is $T_+$ ($T_-$) just in front of (behind) the wall.  The fluid velocity likewise is discontinuous there.  

Since it is often convenient to switch between reference frames, we adopt the notation $v_{xy}$ for the velocity of $x$ in the reference frame $y$. In this context, $x$ and $y$ either refer to the wall ($w$), shock front ($s$), or the fluid at position $1$ (in front of the wall), $2$ (behind the shock front), or $u$ (the unperturbed ``universe'' frame, in front of the shock front or behind the wall). The wall velocity, which is measured with respect to the fluid directly in front of the wall, is  $v_{w1}$ in this notation. We note that for any $x$ and $y$, $v_{xy}=-v_{yx}$. A diagram depicting the geometry and labels is shown in fig.\ \ref{fig:DeflagrationLabels}.

The relationships between the various fluid velocities and temperatures are found by integrating the stress tensor $T_{\mu\nu}$ across either of the two interfaces shown in fig.\ \ref{fig:DeflagrationLabels}. Approximating the
fluid as perfect, these depend only on the fluid density and pressure. The equations of state can be expressed as \cite{Kozaczuk:2015owa}
\begin{equation}\label{eq:EOSpressure}
p_\pm = \frac{1}{3}a_\pm(T_n) T_\pm^4 - \epsilon_\pm(T_n)
\end{equation}
\begin{equation}\label{eq:EOSdensity}
\rho_\pm = a_\pm (T_n) T_\pm^4 + \epsilon_\pm(T_n)
\end{equation}
where $\rho_\pm$ is the fluid density on either side of the bubble wall and $p_\pm$ is the pressure. $a_\pm$ and $\epsilon_\pm$ are given by
\begin{equation}
a_\pm(T) = -\frac{3}{4 T^3}\,\frac{d \mathcal{F}_\pm(T)}{dT}
\end{equation}
\begin{equation}
\epsilon_\pm(T) = \mathcal{F}_\pm(T) + \frac{1}{3}a_\pm(T) T^4 \,.
\end{equation}
$\mathcal{F}_\pm$ is the free energy of the fluid evaluated at the respective VEVs outside and inside wall and $T=T_n$. It is given by
\begin{equation}
\mathcal{F}(h,s,T) = V_{\rm eff}(h,s,T) - \frac{g_*'\, \pi^2}{90}T^4\,.
\end{equation}
Here $g_*'=107.75 - 24.5= 83.25$ is the effective number of degrees of freedom, apart from the $t$,
$W/Z$, $h$, $\chi$ and $s$, whose contributions are already included in $V_{\rm eff}$. In general, $a_\pm(T)$ and $\epsilon_\pm(T)$ in  (\ref{eq:EOSpressure},\ref{eq:EOSdensity}) should be evaluated at the temperatures $T_\pm$, but for transitions typically of interest for baryogenesis, where there is a limited degree of supercooling, the $T$-dependence of $a(T)$ and $\epsilon(T)$ is insignificant.

Integrating $T_{\mu\nu}$ across the bubble wall provides the relations \cite{Espinosa:2010hh}
\begin{equation} \label{eq:wallTeq1}
v_{w1} v_{wu} = \frac{p_+ - p_-}{\rho_+ - \rho_-},\quad
\frac{v_{w1}}{v_{wu}} = \frac{p_+ + \rho_-}{\rho_+ + p_-}\,.
\end{equation}
Integrating across the shock front, the temperature
changes, 
but not the 
the field values, leading to 
\begin{equation} \label{eq:wallTeq3}
v_{su} v_{s2} = \frac{1}{3},\quad
\frac{v_{su}}{v_{s2}} = \frac{T_n^4+3 T_s^4}{3T_n^4 + T_s^4}\, .
\end{equation}
Fluid velocities in the wall and shock wave frame are related by Lorentz transforming to the $u$ frame using
\begin{equation}
v_{2u} = \frac{v_{su} - v_{s2}}{1-v_{su}v_{s2}},\quad
v_{1u} = \frac{v_{wu} - v_{w1}}{1-{v_{wu}v_{w1}}} \,.
\end{equation}

\begin{figure}[htb!]
	\centering
	\includegraphics[width=0.45\textwidth]{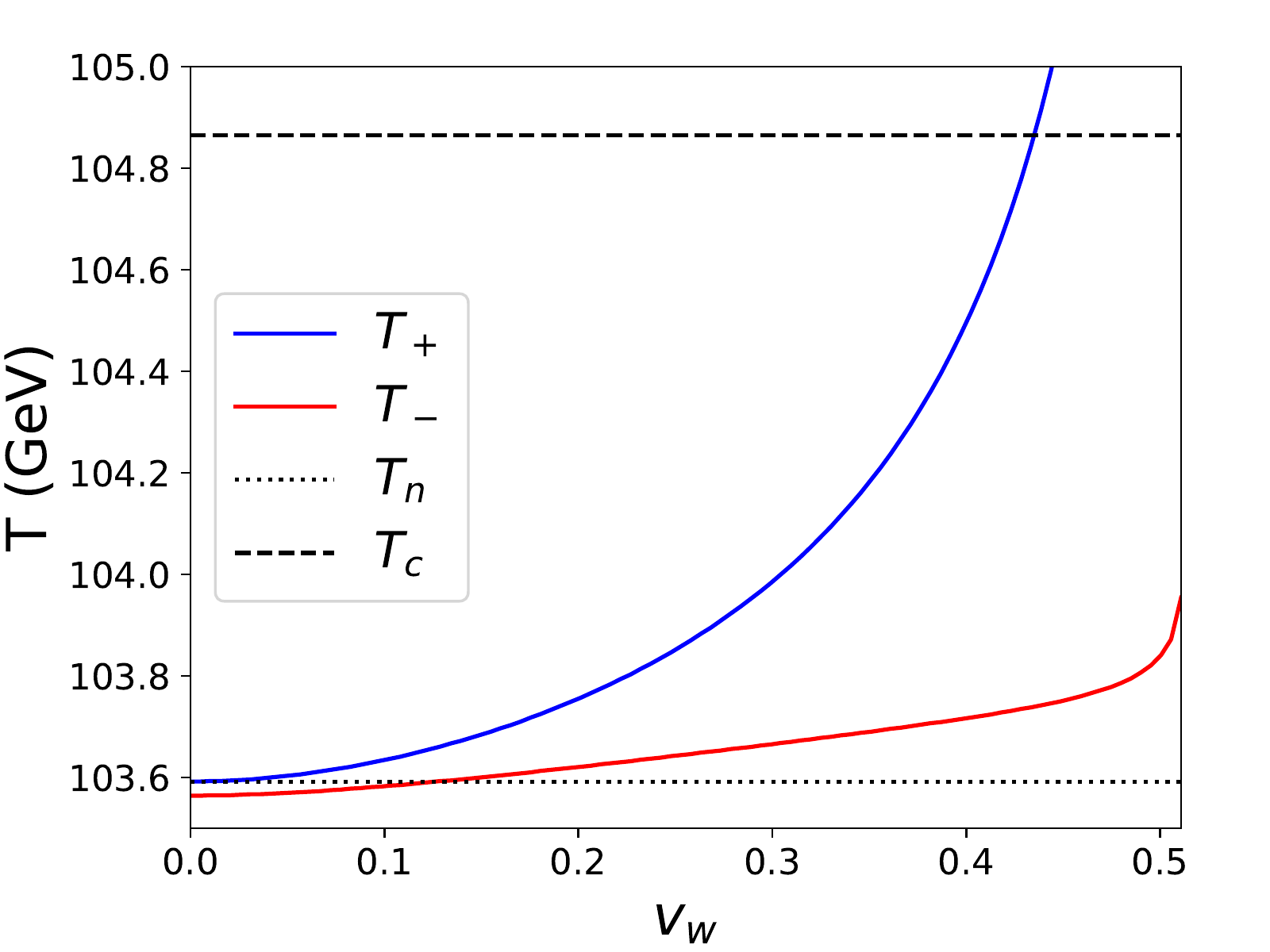}
	\caption{An example of how the wall temperature changes as a function of $v_w$ for a sample model with $m_s = 63\textrm{ GeV}$, $w_0 = 130\textrm{ GeV}$, and $\lambda_{hs} = 0.9$ }
	\label{fig:Tofv}
\end{figure}

The relationship between the fluid velocity and temperature behind the shock front and in front of the wall can be approximated by linearizing
the stress-energy equations
with respect to small fluid velocities in the universe  frame \cite{Huber:2013kj}, to obtain
\begin{equation}\label{eq:wallTeq7}
T_s = T_+ \exp\bigg[ \frac{-2v_{u1}v_{wu}^2}{1-3v_{wu}^2} \bigg(\frac{1}{v_{wu}} - \frac{1}{v_{su}} \bigg) \bigg]
\end{equation}
\begin{equation} \label{eq:wallTeq8}
v_{2u} = v_{1u}\bigg( \frac{v_{wu}}{v_{su}}\bigg)^2 \bigg( \frac{3v_{su}^2 - 1}{3 v_{wu}^2 - 1}\bigg)
\end{equation}
For a given guess of the wall velocity, $v_w \equiv v_{w1}$, and an associated nucleation temperature, eqs.\ (\ref{eq:wallTeq1}-\ref{eq:wallTeq8}) can be used to solve for the eight remaining variables: $T_+$, $T_-$, $T_s$, $v_{wu}$, $v_{su}$, $v_{1u}$, $v_{2u}$, and $v_{s2}$.
An example of the solution to these equations for sample set of parameters is shown in fig.\ \ref{fig:Tofv}.

\subsection{Equations of Motion}
The equation of motion of a scalar field coupled to a perfect fluid has been derived by enforcing the conservation of the stress-energy tensor in the WKB approximation \cite{Moore:1995si}, or starting from the Kadanoff-Baym equations \cite{Konstandin:2014zta}. Both methods lead to
\begin{equation}
    \Box\phi + \frac{\partial V(\phi)}{d\phi} +\sum_i n_i \frac{dm_i^2}{d\phi}\int\frac{d^3p}{(2\pi)^3 2E}\,f_i(\vec{p},x)=0
\end{equation}
where $V(\phi)$ is the zero-temperature effective potential, the sum is over all particles that couple to $\phi$, $n_i$ is the number of degrees of freedom of particle $i$, $m_i$ is its field-dependent mass, and $f_i(\vec{p},x)$ is its phase space distribution. By separating $f_i = f_{0,i} + \delta\! f_i$ into equilibrium and out-of-equilibrium components, the equation of motion takes a more useful form. The integral over $f_{0,i}$ is equivalent to accounting for the $T$-dependence of the effective potential, giving
\begin{equation} \label{eq:scalarEOM}
    \Box\phi + \frac{\partial V_{\rm eff}(\phi,T)}{d\phi} +\sum_i n_i \frac{dm_i^2}{d\phi}\!\!\int\!\!\frac{d^3p}{(2\pi)^3 2E}\,\delta\! f_i(\vec{p},x)=0
\end{equation}
The third term in (\ref{eq:scalarEOM}) describes the friction force that comes from the dissipative interactions between the scalar field and the surrounding fluid. 

In the following, we assume that the dominant sources of friction are the top quark and 
electroweak gauge bosons. The lighter fermions, gluons and photons can be safely ignored because of their negligible couplings to the Higgs field. The self-couplings and mixing of the scalar fields are assumed to be subdominant due to their fewer degrees of freedom relative to vector bosons or quarks.  It is possible that including these contributions would lead to moderately slower walls, which could be advantageous for
baryogenesis, but we defer this issue to future study.

For the electroweak phase transition considered here, there are two relevant scalar fields, each with its own equation of motion, that must be simultaneously solved. Eq.\ (\ref{eq:scalarEOM}) can be further simplified by accounting for the spherical symmetry of the wall and going to the planar limit, which
reduces the system to one spatial dimension, and by considering only the steady-state regime. Therefore, the equations of motion for a bubble wall traveling in the negative $z$ direction become
\bea \label{eq:eom1}
    -h''(z) &+& \frac{\partial V_{\rm eff}(h,s,T)}{\partial h}\\
    &+& \sum_{i=t,W,Z} n_i \frac{dm_i^2}{dh}\int \frac{d^3p}{(2\pi)^3 2E}\,\delta\! f_i(\vec{p},z) = 0\nn\\
    -s''(z) &+& \frac{\partial V_{\rm eff}(h,s,T)}{\partial s} = 0
\eea
where primes denote derivatives with respect to $z$.
Strictly speaking, the existence of a steady-state solution to the equations of motion does not
guarantee that those solutions will in fact be realized in the physical setting, but this issue
is beyond the scope of this paper.

\subsection{Friction}
The friction experienced by the wall depends on $\delta\! f_i(\vec p,z)$, the deviation from equilibrium of $W/Z$
and $t$. We adopt the fluid approximation framework developed in \cite{Moore:1995si}, in which the friction is fully described by three fluids: that of the top quark, the massive gauge bosons, and the other particles, denoted as the ``background.''\ \  We label the gauge
boson contribution by $W$ although it also includes $Z$.
For simplicity $W$ and $Z$ are grouped together due to their similar couplings, and assigned a mass-squared that is the weighted average of $m_W^2$ and $m_Z^2$. The background fluid encompasses all the fields that are assumed to contribute negligible friction, but which nevertheless
play an important role in the wall dynamics. We consider
friction only from fluid excitations with large momentum, such that the wavelength is shorter than the width of the wall. It has been shown that IR excitations in the massive gauge boson fluid can be important \cite{Moore:2000wx}, but we have checked numerically that
these are subdominant for parameters of interest in 
the present study, using the same approximations to evaluate the IR contributions as in ref.\ \cite{Kozaczuk:2015owa},
but taking care to impose the perturbative cutoff 
$m_W(z) > g^2 T$ \cite{Moore:2000wx}.

The phase space distribution for the $t$ and $W$ fluids can be parametrized as
\begin{equation} \label{eq:fiDef}
f_i(E, z) = \frac{1}{e^{(E + \delta_i(z))/T} \pm 1}
\end{equation}
where the $+/-$ is for fermions/bosons and 
\bea \label{eq:pertDef}
\delta_i(z) &=& -\Big[T(\delta\mu_i + \delta\mu_{bg})(z) + E(\delta\tau_i + \delta\tau_{bg})(z)\nn\\ &+& p_z(\delta v_i + \delta v_{bg})(z)\Big]
\eea
accounts for perturbations in the fluids. $\delta\mu_i(z)$, $\delta\tau_i(z)$, $\delta v_i(z)$
are respectively the perturbations in the chemical potential, the relative temperature and the velocity. 
The  subscript $bg$ denotes the background fluid. All the perturbations are relative to the fluid directly in front of the wall where $\mu =0$ and $T=T_+$ as described in section \ref{sec:Temp}.


\begin{figure*}[htb!]
	\centering
	\includegraphics[width=\textwidth]{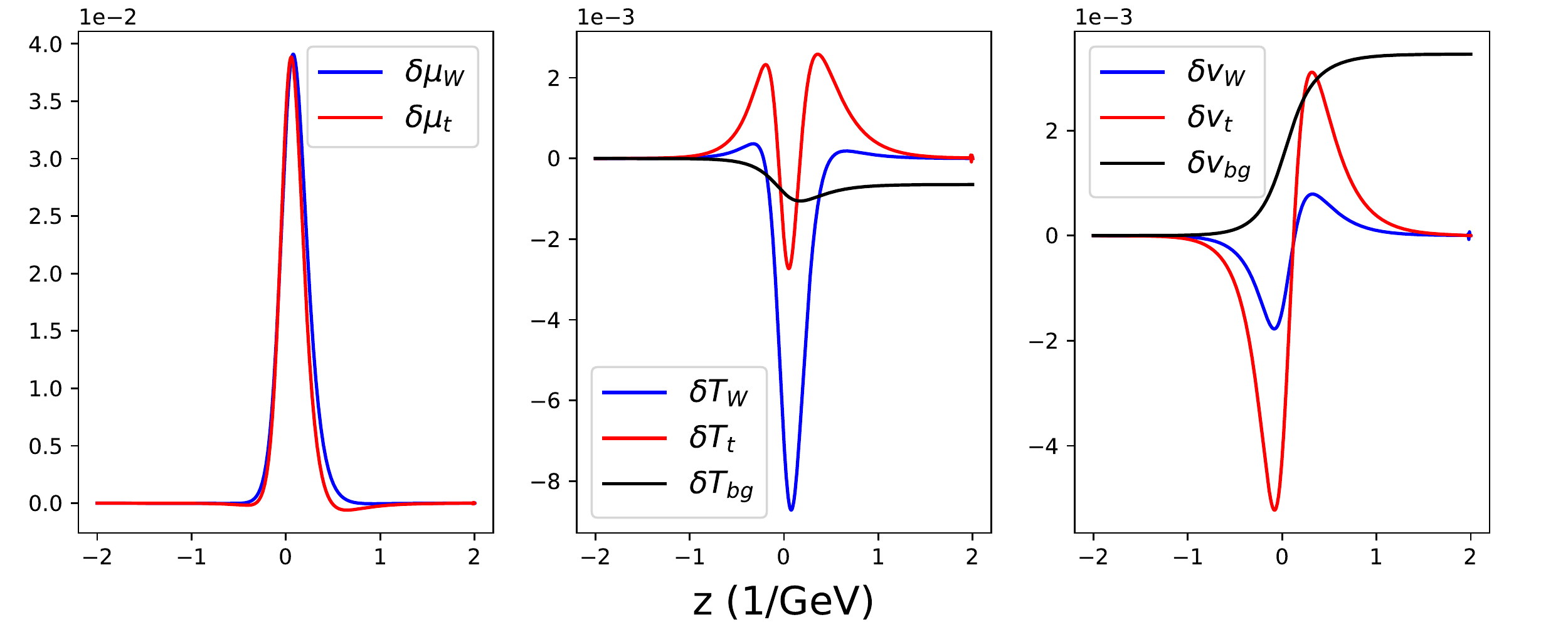}
	\caption{Example of the solutions for the fluid perturbations, for a model with $m_s = 63\textrm{ GeV}$, $w_0 = 130\textrm{ GeV}$, and $\lambda_{hs} = 0.9$, $v_w = 0.128$ and background wall shape $h(z) = ({h_0}/{2})[\tanh({z}/{L_w}) + 1]$ where $h_0 = 209\textrm{ GeV}$ and $L_w = 0.2\textrm{ GeV}^{-1}$. }
	\label{fig:perts}
\end{figure*}

Deviations from equilibrium in the fluids are governed by the Boltzmann equation
\begin{equation} \label{eq:boltzEq}
{d\over dt} f_i(E,z) = -C[f_i(E,z)]\,.
\end{equation}
Rather than solving 
the full Boltzmann equation, one linearizes it in  $\delta_i(z)$, and converts it to a system of ordinary differential equations by taking three moments:  $\int {d^{\,3}p}/{(2\pi)^3}$, $\int (E/T)\,{d^{\,3}p}/(2\pi)^3$, and $\int{p_z\,d^{\,3}p}/{(2\pi)^3}$. A  detailed derivation is provided in appendix \ref{sec:AppendBoltz}, leading to the coupled
matrix equations
\bea \label{eq:linBoltzW}
 A_\W (\vec{q}_\W + \vec{q}_{bg})' + \Gamma_\W \vec{q}_\W &=& S_\W\\
 \label{eq:linBoltzt}
 A_t (\vec{q}_t + \vec{q}_{bg})' + \Gamma_t \vec{q}_t &=& S_t\\
\label{eq:linBoltzbg}
 A_{bg} \vec{q}^{\,\prime}_{bg} + \Gamma_{bg,\W} \vec{q}_\W + \Gamma_{bg,t} \vec{q}_t&=& 0
\eea
where  $\vec{q}_i^{\textrm{ }T}=~(\delta\mu_i, \delta\tau_i, \delta v_i)$. The $A_i$ matrices for $i=W,t$ take the form
\begin{equation} \label{eq:ADef}
A_i \equiv
\begin{bmatrix}
v_w c_2^i & v_w c_3^i & \frac{1}{3}d_3^i \\
v_w c_3^i & v_w c_4^i & \frac{1}{3}d_4^i \\
\frac{1}{3}d_3^i & \frac{1}{3}d_4^i & \frac{1}{3}v_w d_4^i
\end{bmatrix}
\end{equation}
while the source terms are 
\begin{equation}
S_i \equiv \frac{m_i' m_i^{\phantom{'}}}{T^2}
\begin{bmatrix}
v_w c_1^i \\
v_w c_2^i \\
0
\end{bmatrix}
\end{equation}

The coefficients $c_j^i$ and $d_j^i$ denote the integrals
\begin{equation}
c_j^i\left(\frac{m_i}{T}\right) \equiv \int \frac{d^3{p}}{(2\pi)^3}\big(-f_{0,i}'\big)\frac{E^{j-2}}{T^{j+1}}
\end{equation}
and
\begin{equation}
d_j^i\left(\frac{m_i}{T}\right) \equiv \int \frac{d^3{p}}{(2\pi)^3}\big(-f_{0,i}'\big)\frac{p^2E^{j-4}}{T^{j+1}}
\end{equation}
where $f_{0,i}$ is the equilibrium distribution function for particle $i$. In previous literature (with the exception of ref.\ \cite{Laurent:2020gpg}),
these coefficients were evaluated in the massless approximation, where $d_j^i=c_j^i$,
setting $m_i/T=0$. But for some phase transitions satisfying the sphaleron bound (\ref{sph_bound}), $m_t/T > 1$ in the broken phase; hence we 
we derived the full
mass dependence for this work. For the background fluid, one can show that 
\begin{equation}
    A_{bg}=20 A_\W|_{m=0}+78 A_t|_{m=0}
\end{equation} where the background fluids are approximated as massless \cite{Konstandin:2014zta}.

\subsubsection{Collision terms} \label{sec:CollisionTerm}

The $\Gamma_i$ matrices in eqs.\ (\ref{eq:linBoltzW},\,\ref{eq:linBoltzt}) quantify the interactions of the fluids. $\Gamma_t$ and $\Gamma_\W$ take the form
\begin{equation}
\Gamma_i \equiv T
\begin{bmatrix}
\Gamma_{\mu 1i} & \Gamma_{\delta T1i} & 0 \\
\Gamma_{\mu 2i} & \Gamma_{\delta T2i} & 0 \\
0 & 0 & \Gamma_{vi}\, .
\end{bmatrix}
\end{equation}
The matrix elements were originally computed in ref.\ \cite{Moore:1995si}, to leading-log
accuracy in the masses of particles exchanged in the various interactions.  This means that the infrared divergence that would arise from $t$-channel exchange of massless particles (in the electroweak symmetric phase) is cut off by taking account of their thermal masses in the propagator, while neglecting such mass effects otherwise.   Several calculational errors in 
\cite{Moore:1995si} were subsequently corrected in Ref.\ \cite{Arnold:2000dr}, which we take into account here.

In ref.\ \cite{Kozaczuk:2015owa}, a refined leading log calculation was carried out, including hard thermal loops, for the top quark and Higgs scattering rates, but not for $W$ bosons, since it was argued there that the $W$ contribution to friction could be determined from IR-dominated modes described by treating the $W$ as a classical field \cite{Moore:2000mx}.   We have found that these contributions are numerically smaller in the present model than the perturbative ones from the original calculation \cite{Moore:1995si}, so this approach would not be consistent here.  Moreover the inclusion of extra annihilation channels like $t\bar t\to hh$, carried out in \cite{Kozaczuk:2015owa}, would not be consistent in our present approach, where we approximate
the Higgs fluid as maintaining thermal equilibrium, by omitting its perturbations from the Boltzmann network.

Instead, we have improved on the original estimates 
of \cite{Moore:2000mx} using results from ref.\ \cite{Laurent:2020gpg}, which recomputed the phase space integrals numerically instead of approximating them analytically\footnote{The significant difference between the numerically evaluated phase-space integrals and the values obtained using the approximations of ref. \cite{Moore:2000mx} was pointed out in ref. \cite{Kozaczuk:2015owa}.}, leading to\footnote{The third row entries differ from those of Ref.\ \cite{Laurent:2020gpg} because a different set of fluid equations were solved in that reference.  We thank B.\ Laurent for recomputing the third rows for the fluid equations used in the present work.}
\bea \label{eq:intRate}
\Gamma_\W/T &=&
\begin{bmatrix}
0.00239 & 0.00512 & 0\\
0.00512 & 0.0174 & 0 \\
0 & 0 & 0.00663
\end{bmatrix}\nn\\
\Gamma_t/T &=& 
\begin{bmatrix}
0.00196 & 0.00445 & 0\\\
0.00445 & 0.0177 & 0\\
0 & 0 & 0.00992
\end{bmatrix}
\eea
The effect of using this collision term compared to that of \cite{Moore:1995si}, is explored in section \ref{sec:vwResults}.

Using energy-momentum conservation, the background fluid collision terms are given by
\begin{equation}
    \Gamma_{bg,i} = -n_i \Gamma_i
\end{equation}
where $n_t=12$ and $n_W = 9$ are the number of degrees of freedom in the respective components.

The background fluid is assumed to be in chemical equilibrium, implying that $\delta \mu_{\rm bg} = 0$. This assumption removes the top row of eq.\ (\ref{eq:linBoltzbg}). The remaining two rows determine
$\delta \tau_{\rm bg}$ and $\delta v_{\rm bg}$ in terms of
$q_\W$ and $q_t$, 
\begin{equation} \label{eq:qbg}
\vec{q}_\textrm{bg}^{\textrm{ }\prime} = -A_{bg23}^{-1}(\Gamma_{\textrm{bg},\W}\vec{q}_\W +\Gamma_{\textrm{bg},t}\vec{q}_t)
\end{equation}
where $A_{bg23}^{-1}$ denotes the matrix 
where the bottom right block of $A_{bg}$ is inverted and the rest of the matrix elements are zero.

Equations (\ref{eq:linBoltzW}) and (\ref{eq:linBoltzt}) can then be expressed in $6\times 6$ matrix form, in the rest frame of the bubble wall,
\begin{equation} \label{eq:WtPert}
A \vec{q}^{\textrm{ } \prime}+\Gamma \vec{q} = S
\end{equation}
with
\bea
A&\equiv& \gamma \begin{bmatrix}
A_W&0\\
0&A_t
\end{bmatrix},\ 
\vec{q} \equiv \begin{bmatrix}
\vec{q}_\W\\
\vec{q}_t
\end{bmatrix},\ 
S \equiv \gamma \begin{bmatrix}
S_\W\\
S_t
\end{bmatrix}
\eea
and
\begin{equation} \label{eq:GamDef}
\Gamma \equiv
\begin{bmatrix}
\Gamma_\W& 0\\
0 & \Gamma_t
\end{bmatrix}
-
\begin{bmatrix}
A_\W A_{\textrm{bg}23}^{-1}\Gamma_{\textrm{bg},\W}& A_\W A_{\textrm{bg}23}^{-1}\Gamma_{\textrm{bg},t}\\
A_t A_{\textrm{bg}23}^{-1}\Gamma_{\textrm{bg},\W}& A_t A_{\textrm{bg}23}^{-1}\Gamma_{\textrm{bg},t}
\end{bmatrix}
\end{equation}
The factors of $\gamma=1/\sqrt{1-v_w^2}$ are from Lorentz boosting to the rest frame of the wall.

The $W$ and $t$ fluid perturbations are determined by solving eq.\ (\ref{eq:WtPert}) using the relaxation method as described in ref.\ \cite{Press:1992}, since shooting tends to be unstable. The background fluid perturbations are found by integrating eq.\ (\ref{eq:qbg}). One can carry out this procedure for
given values of the wall velocity and shape, and from 
the ensuing perturbations compute the friction term in the Higgs field equation of motion (\ref{eq:eom1}) using
\bea\label{eq:friction}
\int\frac{d^3p}{(2\pi)^3 2E}\,\delta\! f_i(\vec{p},z) &\cong& \int\frac{d^3p}{(2\pi)^3 2E}f_{0,i}'(\vec{p},z)\,\delta_i(z) \nn \\
=\frac{T^2}{2}\Big[c_1^i(z)\delta\mu_i(z) &+& c_2^i(z)\big(\delta\tau_i(z)+\delta\tau_{bg}(z)\big)\Big]\nn\\
\eea
An example of the solutions for the perturbations is shown in fig.\ \ref{fig:perts}.

\begin{figure}[b]
	\centering
	\includegraphics[width=0.45\textwidth]{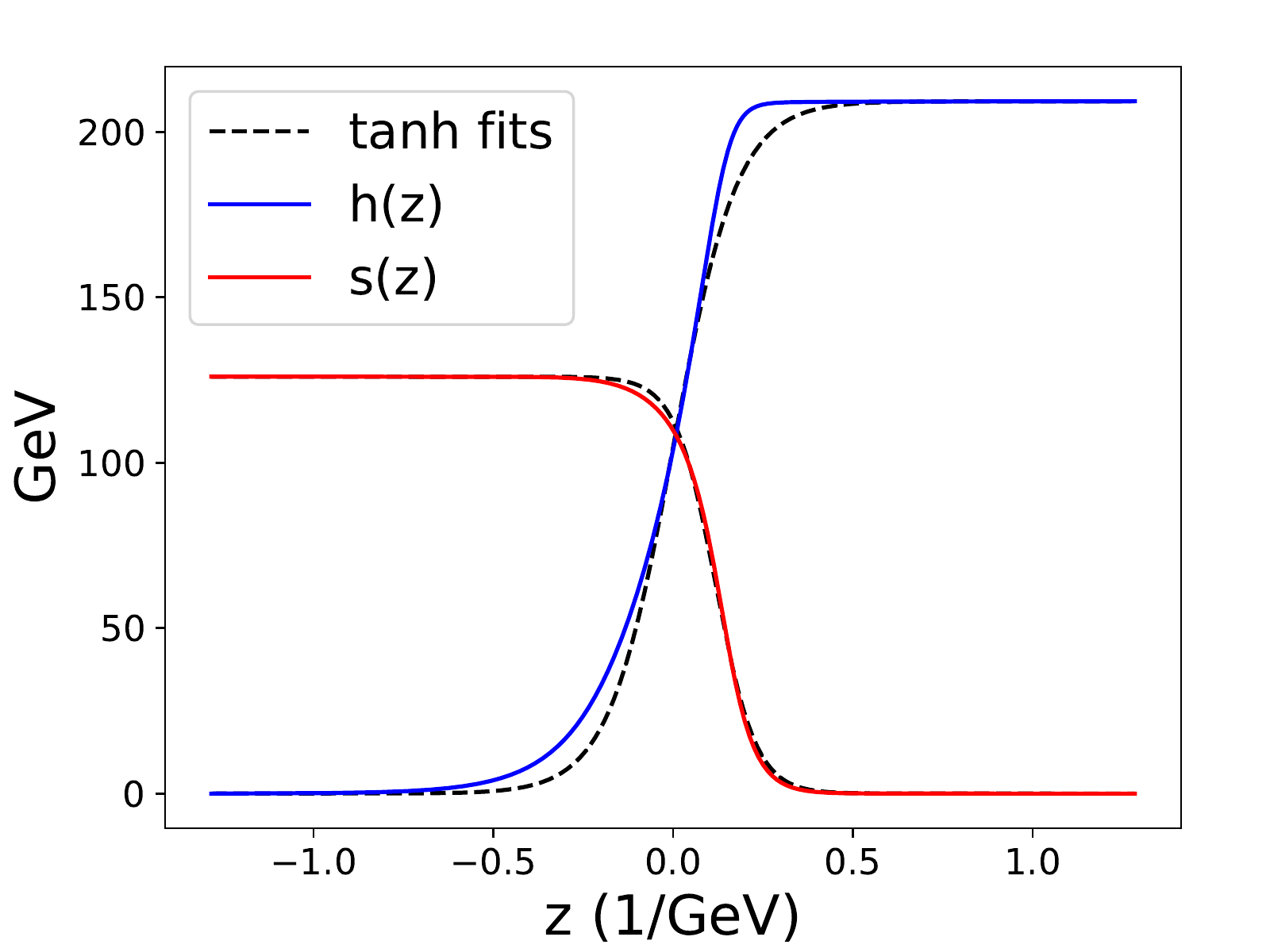}
	\caption{The wall shape that solves the equation of motion for the model with $m_s = 63\textrm{ GeV}$, $w_0 = 130\textrm{ GeV}$, and $\lambda_{hs} = 0.9$. The dashed curves show the best fits using the tanh ansatz of eqs.\ (\ref{eq:hfit}~,~\ref{eq:sfit}).}
	\label{fig:wallShape}
\end{figure}

\subsection{Solving the Equations of Motion}
With the friction calculated in eq.\ (\ref{eq:friction}), the equations of motion that must be solved to determine $v_w$ and the shape of the wall are
\bea \label{eq:eom2}
-h''(z) &+& \frac{\partial V_\textrm{eff}(h,s,T_+)}{\partial h}\\
    &+& \frac{n_t T_+}{2} \frac{dm_t^2}{dh}\left[c_1^t\delta\mu_t + c_2^t(\delta\tau_t+\delta\tau_{bg})\right]\nn\\
    &+& \frac{n_W T_+}{2} \frac{dm_W^2}{dh}\left[c_1^W\delta\mu_W + c_2^W(\delta\tau_W+\delta\tau_{bg})\right] = 0\nn\\
 -s''(z) &+& \frac{\partial V_\textrm{eff}(h,s,T_+)}{\partial s} = 0\,.
\eea

Deep into the bubble interior, eq.\ (\ref{eq:eom2}) is not exactly satisfied once we adopt our approximation schemes for calculating the effective potential and the perturbations. The Higgs' VEV is unchanging there, so the kinetic term is zero. Similarly the perturbations in the $W$ and $t$ fluids go to zero on both sides of the wall. This implies that the terms proportional to $\delta \tau_{bg}$ must exactly cancel out the potential term. When the perturbations are determined as described above and the potential term is calculated with the Higgs VEV that minimizes the potential inside the bubble, the two terms do not cancel as they should. This is due to differences in the derivation of the friction terms in comparison to the effective potential. Firstly, the fluid perturbations are only determined to linear order whereas the temperatures that go into the effective potential, $T_+$ and $T_-$, were calculated including non-linearities in the fluid equations. This means that while in theory $T_+ - T_- = T_+ \delta\tau_{bg}$, their relationship is only approximate. The other cause is that the scalar fields were treated as massless background fields in the friction calculation but their full contribution was included in the effective potential. There are three ways to account for this inconsistency: the Higgs VEV inside the bubble can be chosen not to minimize the potential but instead to cancel the friction term, the entire friction can be scaled to cancel the potential term but maintaining the friction shape in $z$, or just the background perturbation contribution to the friction can be scaled to cancel the potential term. We adopt the last option, which we found to be the most conservative choice (leading to slightly larger wall velocities).  
The equations of motion that we actually use to determine the wall dynamics then become
\bea \label{eq:eom3}
 E_h&\equiv&   -h''(z) + \frac{\partial V_\textrm{eff}(h,s,T_+)}{\partial h}\\
    &+& \frac{n_t T_+}{2} \frac{dm_t^2}{dh}\left[c_1^t\delta\mu_t + c_2^t(\delta\tau_t+y\delta\tau_{bg})\right]\nn\\
    &+& \frac{n_W T_+}{2} \frac{dm_W^2}{dh}\left[c_1^W\delta\mu_W + c_2^W(\delta\tau_W+y\delta\tau_{bg})\right] = 0\nn\\
  E_s &\equiv&  -s''(z) + \frac{\partial V_\textrm{eff}(h,s,T_+)}{\partial s} = 0\,
\eea
where $y$ is an $\mathcal{O}(1)$ parameter chosen so that the equations are satisfied for larger positive values of $z$.

For a given value of $v_w$, the relaxation method can be used to find the shapes of $h(z)$ and $s(z)$ that come closest to solving the equations of motion. One must then
vary $v_w$ and find a complete solution to the equations, by
iterating this procedure.  
A reasonable initial guess for both $v_w$ and the wall shape
is required, leading us to solve the equations in two stages.
The first part is to guess $v_w$ and the wall shape using the $\tanh$ ansatz employed in previous studies of wall velocities \cite{Moore:1995si},\cite{Konstandin:2014zta},\cite{Kozaczuk:2015owa}. The second uses these as a starting point to 
numerically determine $v_w$ and the wall shapes.

The $\tanh$ ansatz in the first stage assumes that the Higgs profile has the form
\begin{equation} \label{eq:tanh}
    h(z) = \frac{v(T_-)}{2}\left(\tanh\left(\frac{z}{L_w}\right) + 1\right)
\end{equation}
where $v(T_-)$ is the Higgs VEV at temperature $T_-$ and $L_w$ is the width of the wall. The friction and shape of the singlet profile are independent of each other, so there is no need to impose a $\tanh$ ansatz for $s$; rather its
profile is found by numerically solving its equation of motion. This reduces the problem to finding values of $v_w$ and $L_w$ that come closest to solving the Higgs equation of motion\footnote{In an alternative implementation of this initial stage, which is also effective, we fix the path through field space as an arc passing through the saddle point,  and we work with the field equation along that path rather than giving priority to $h$ or $s$.}. 

No choice of $v_w$ and $L_w$ will exactly solve
eq.\ (\ref{eq:eom3}), since the true shape is not a $\tanh$ function.
Instead we follow ref.\ \cite{Konstandin:2014zta} by calculating two moments of $E_h$ in 
eq.\ (\ref{eq:eom3})  and finding the values of $v_w$ and $L_w$
that make them vanish.  The two moments are
taken to be
\bea
    E_1 &\equiv& \int h'(z)\,E_h\,dz= 0\\
    E_2 &\equiv& \int h'(z)\big(2h(z)-v(T_-)\big)\,E_h\,dz= 0\nn
\eea
since with this choice the Jacobian matrix $\partial(E_1,E_2)/\partial(v_w,L_w)$ is always far from being singular.

The first stage of the algorithm can then be summarized as:
\begin{enumerate}
    \item Make a guess for $v_w$ and $L_w$
    \item Calculate $T_+$ and $T_-$ for $v_w$
    \item Determine $s(z)$ by solving the $s$ equation of motion using the $\tanh$ ansatz for $h(z)$
    \item Determine the shape of the friction term for the guessed shape of $h(z)$
    \item Calculate the moments $E_1$ and $E_2$
    \item Find the new guess for $v_w$ and $L_w$ by solving $E_i=0$.
\end{enumerate}

\begin{figure*}[htb]
	\centering
	\includegraphics[width=\textwidth]{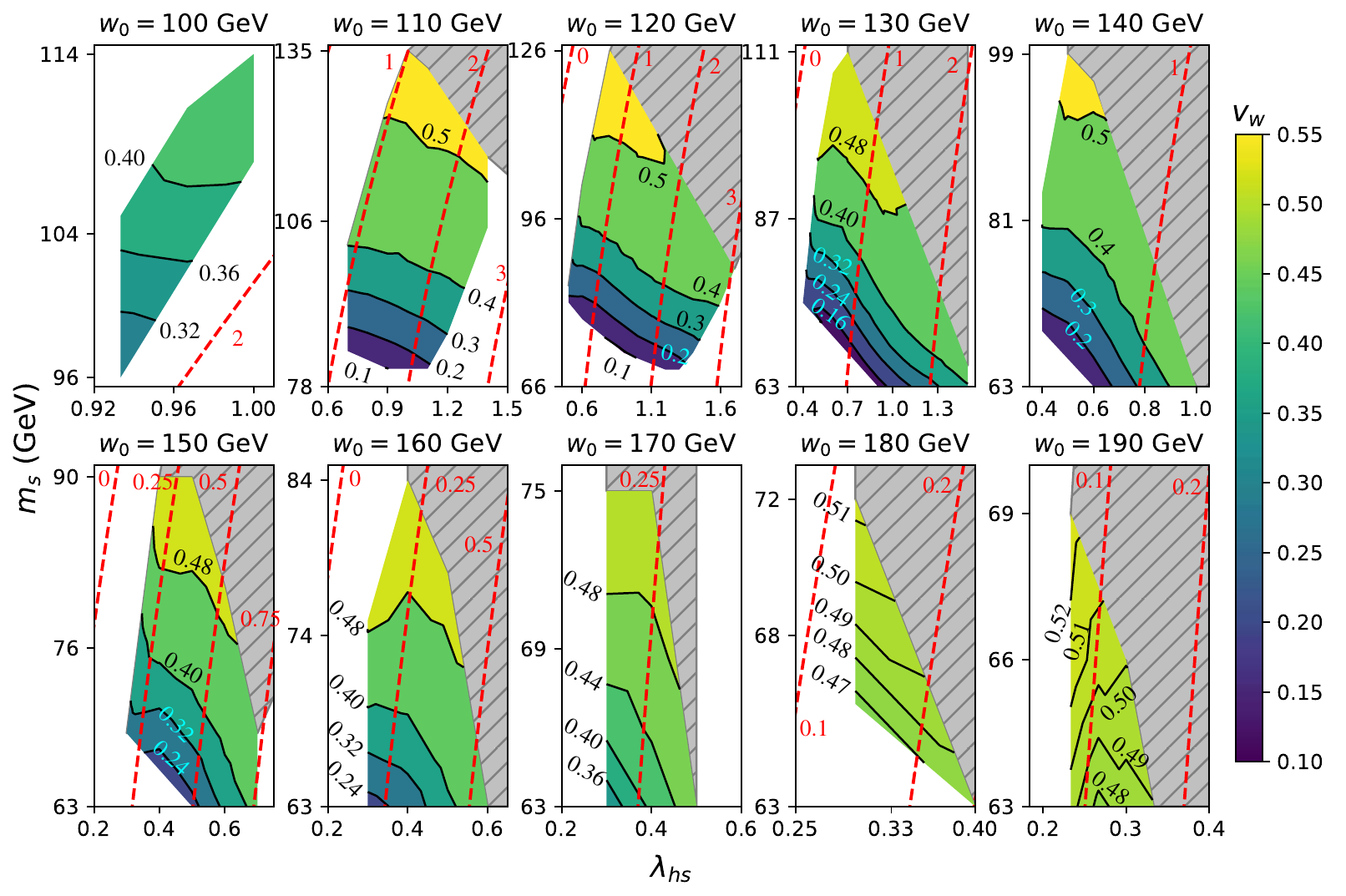}
	\caption{Contours of the wall velocity $v_w$
	in the $\lambda_{hs}$-$m_s$ plane, with $w_0$
	increasing from 100 to 190 GeV in successive plots.
	The white area indicates regions where no first order transition satisfying the sphaleron bound (\ref{sph_bound}) was found. In the grey hatched region, strong transitions satisfying (\ref{sph_bound})
	exist, but no solutions with $v_w<c_s$ were found. The red dashed contours indicate values of the singlet self-coupling, $\lambda_s$, as determined by eq.\ (\ref{smass}). For each $w_0$ we show only regions	containing viable solutions for the bubble wall parameters, within the ranges specified in Eqn.~(\ref{scanregion}).
	}
	\label{fig:scan}
\end{figure*}

In the second stage, we aimed to relax the $\tanh$ profile assumption for $h(z)$ and to determine its shape more exactly.  Using the values of $v_w$ and $L_w$ from the first part as new initial guesses, we solved both $h$ and $s$ 
equations of motion simultaneously, using relaxation.  A challenge here is that
the friction on the wall, which is expensive to compute, depends on the background $h(z)$ solution.  To speed up the algorithm, we recomputed the friction only after several
relaxation steps.  This procedure leads to eventual convergence, unless the initial guess for $v_w$ is too poor.
Convergence was tested by seeing how closely the two equations of motion were satisfied, using the squared error statistic
\begin{equation}
    E_{\rm tot} = \int\left[E_h^2 + E_s^2\right]dz\,.
\end{equation}
The best value of $v_w$ was determined by varying $v_w$ in the region of the guess from step 1 as to minimize $E_{\rm tot}$. An example of the wall shapes that solve the equations of motion is given in fig.\ \ref{fig:wallShape}.  It demonstrates that the actual profiles can differ significantly from the tanh ansatz.

\begin{figure}[t]
	\centering
	\includegraphics[width=0.48\textwidth]{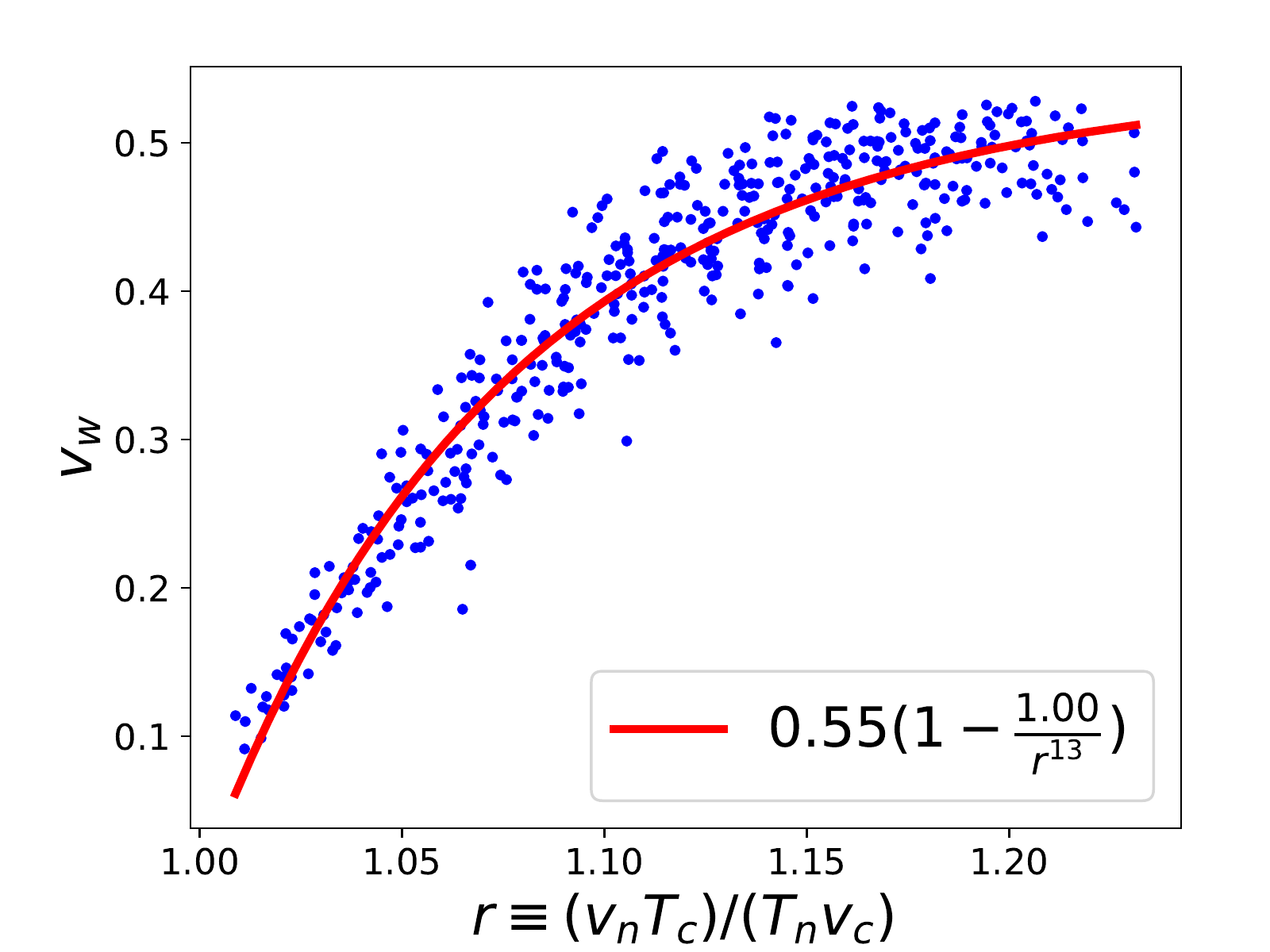}
	\caption{The dependence of wall velocity, $v_w$, on the supercooling parameter, $r = (v_n/T_n)/(v_c/T_c)$. The solid curve shows a fit to the points.}
	\label{fig:supercooling}
\end{figure}

\begin{figure}[t]
	\centering
	\includegraphics[width=0.48\textwidth]{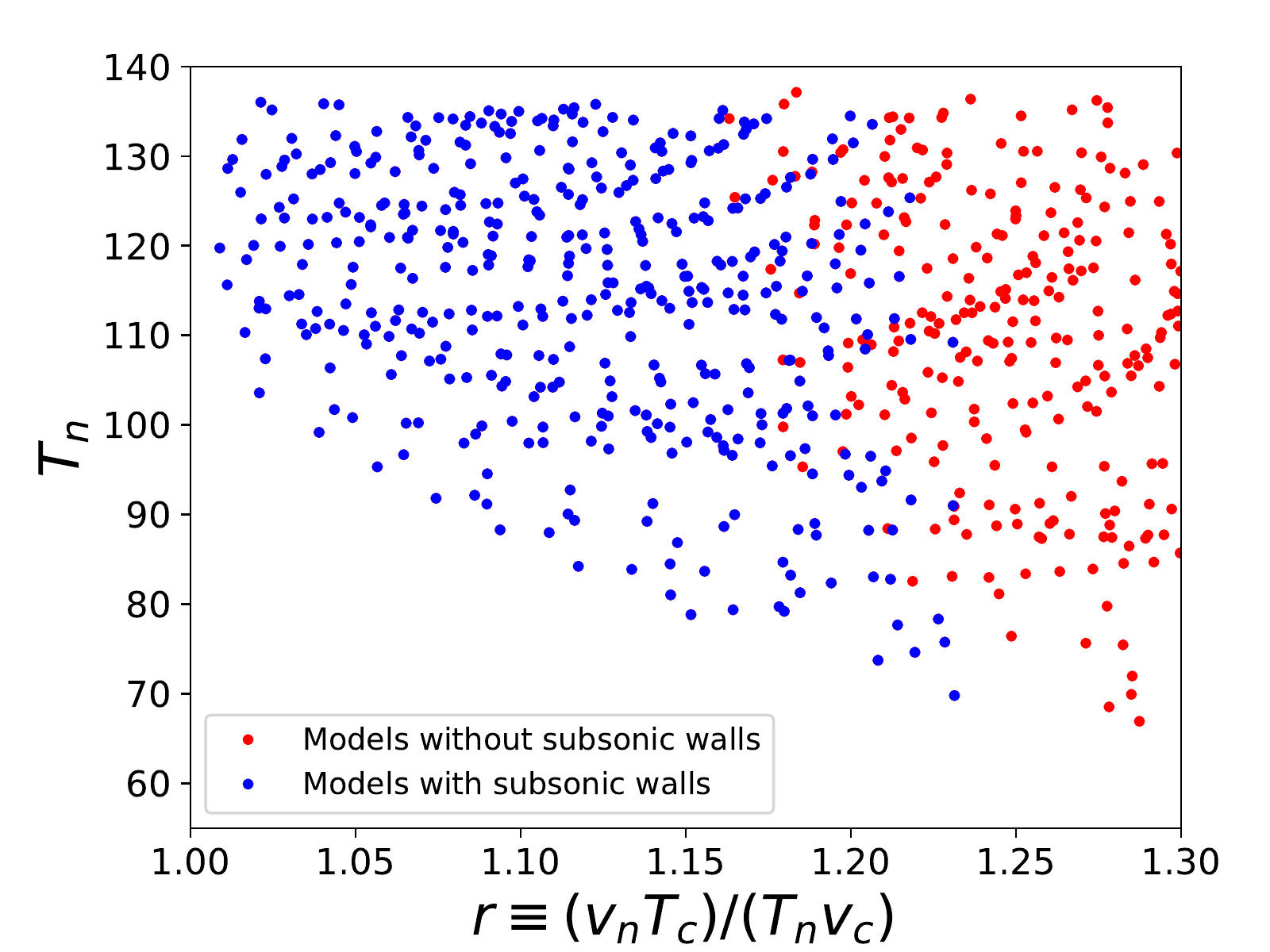}
	\caption{Scatter plot of nucleation temperature $T_n$ versus 
	the supercooling parameter $r$,
	for all models CosmoTransitions found to satisfy the sphaleron washout condition.}
	\label{fig:rvsTn}
\end{figure}

\begin{figure}[t]
	\centering
	\includegraphics[width=0.48\textwidth]{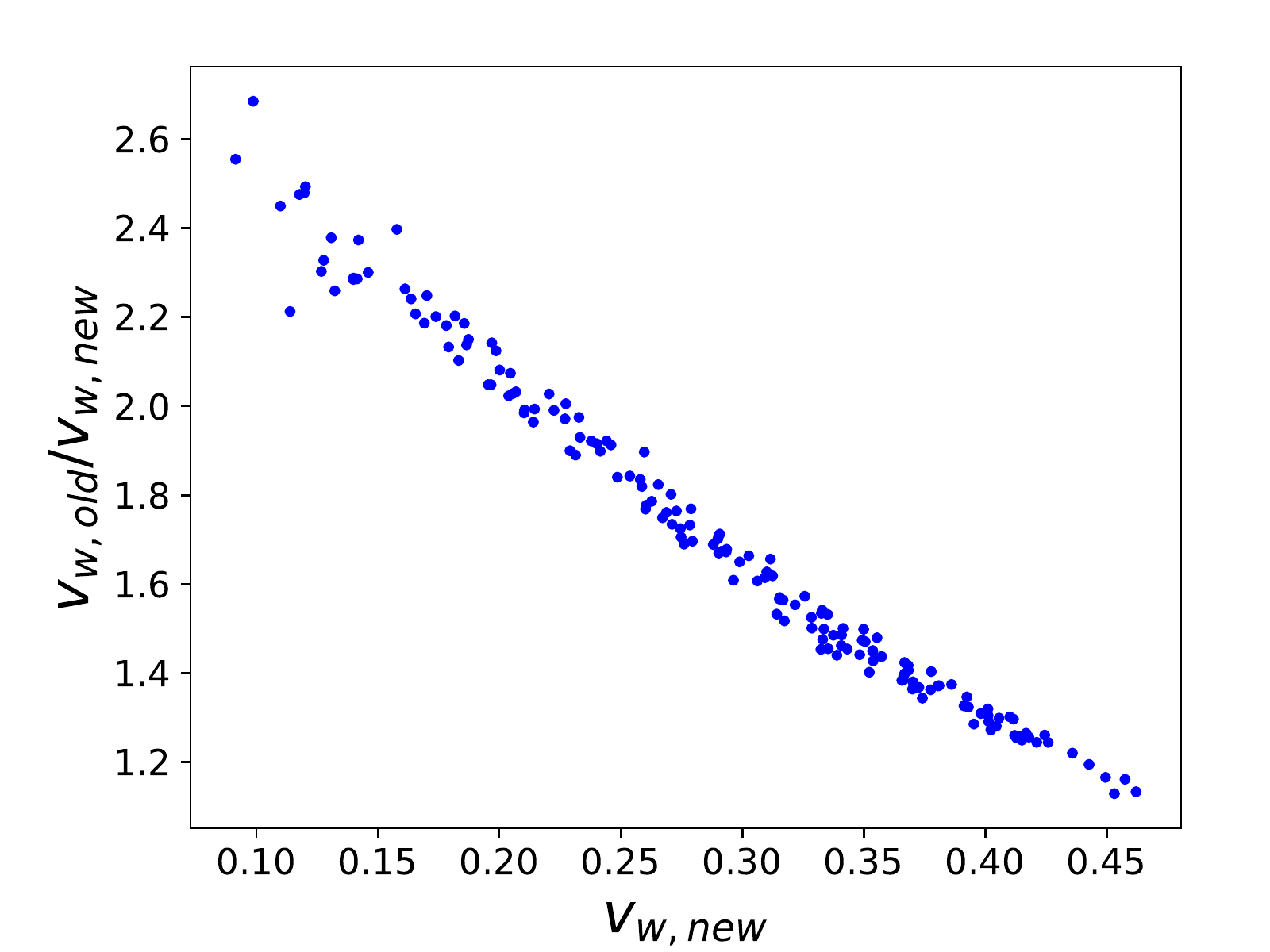}
	\caption{Comparison of wall velocities determined using the corrected interaction rates in eq. \eqref{eq:intRate} ($v_{w,new}$) compared to the incorrect wall velocities that would be found by using the interaction rates reported in \cite{Moore:1995si} ($v_{w,old}$).}
	\label{fig:intRateComparison}
\end{figure}

\begin{figure}[t]
	\centering
	\includegraphics[width=0.48\textwidth]{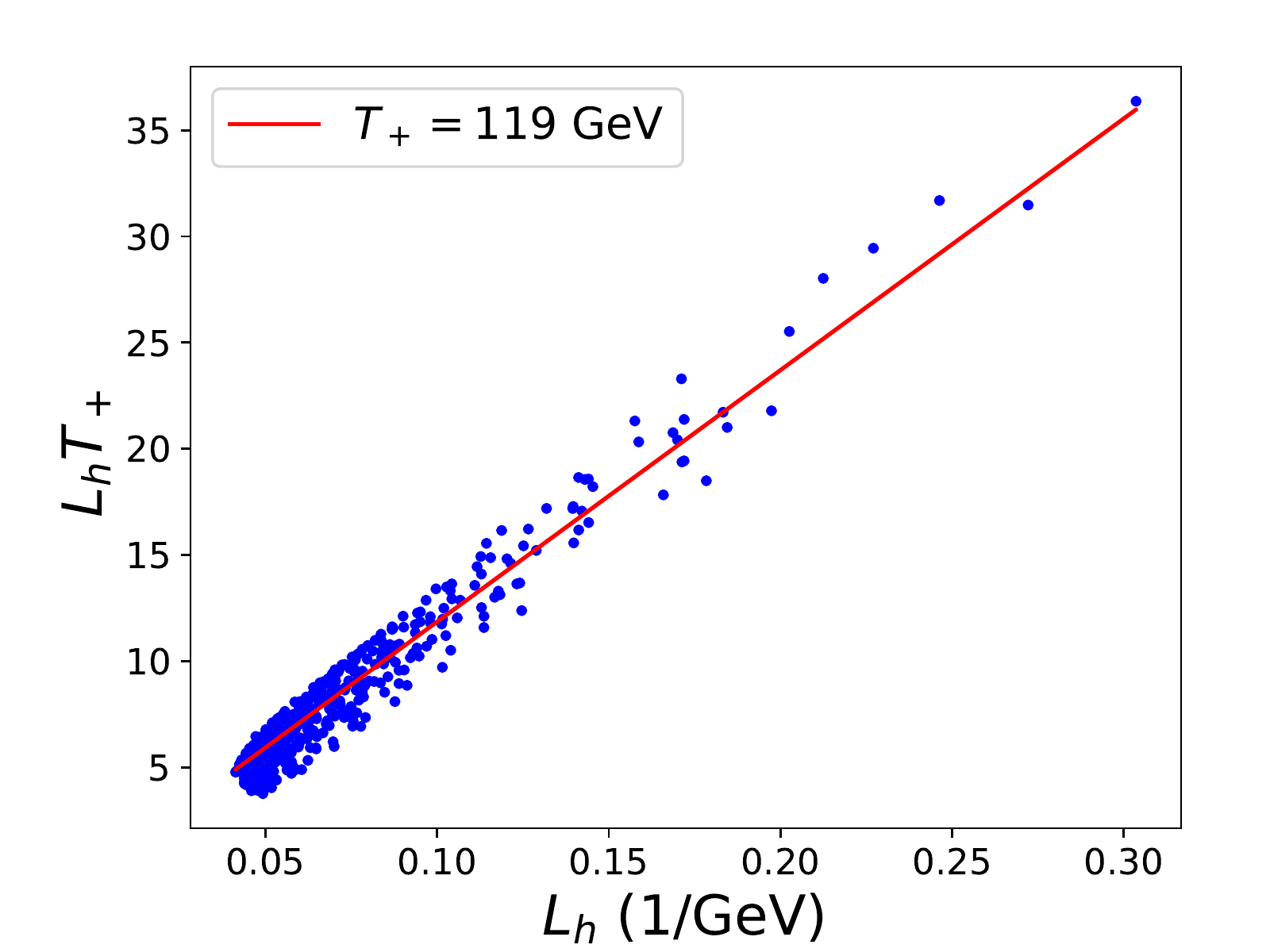}
	\caption{The Higgs wall width $L_h$, in units of the inverse wall temperature $T_+^{-1}$, versus the same quantity in GeV$^{-1}$ units. The solid line shows a fit to the points, corresponding to the mean wall temperature $T_+=119$\,GeV.}
	\label{fig:LvsLT}
\end{figure}

\begin{figure*}[t]
	\centering
	\includegraphics[width=0.95\textwidth]{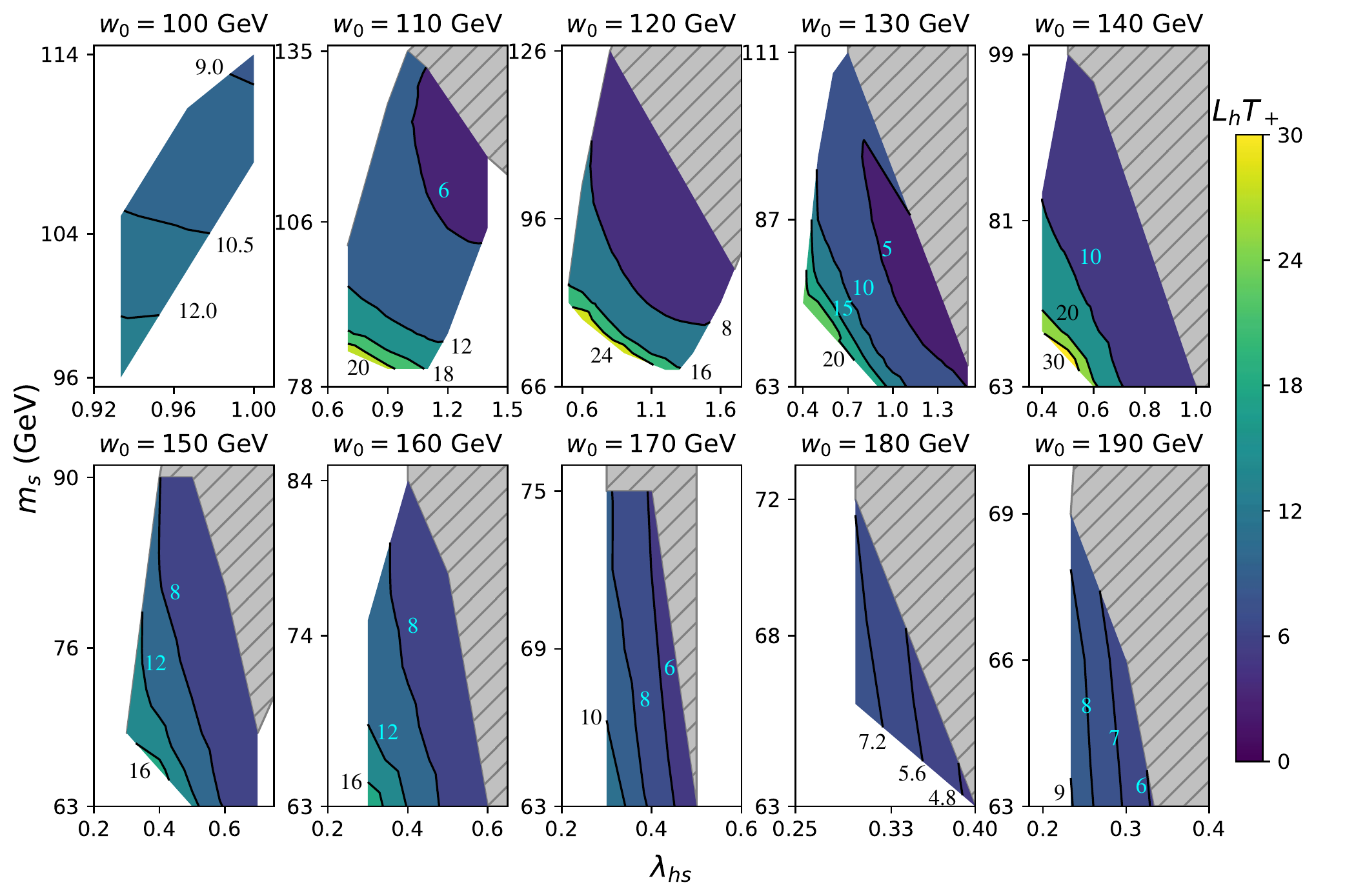}
	\includegraphics[width=0.95\textwidth]{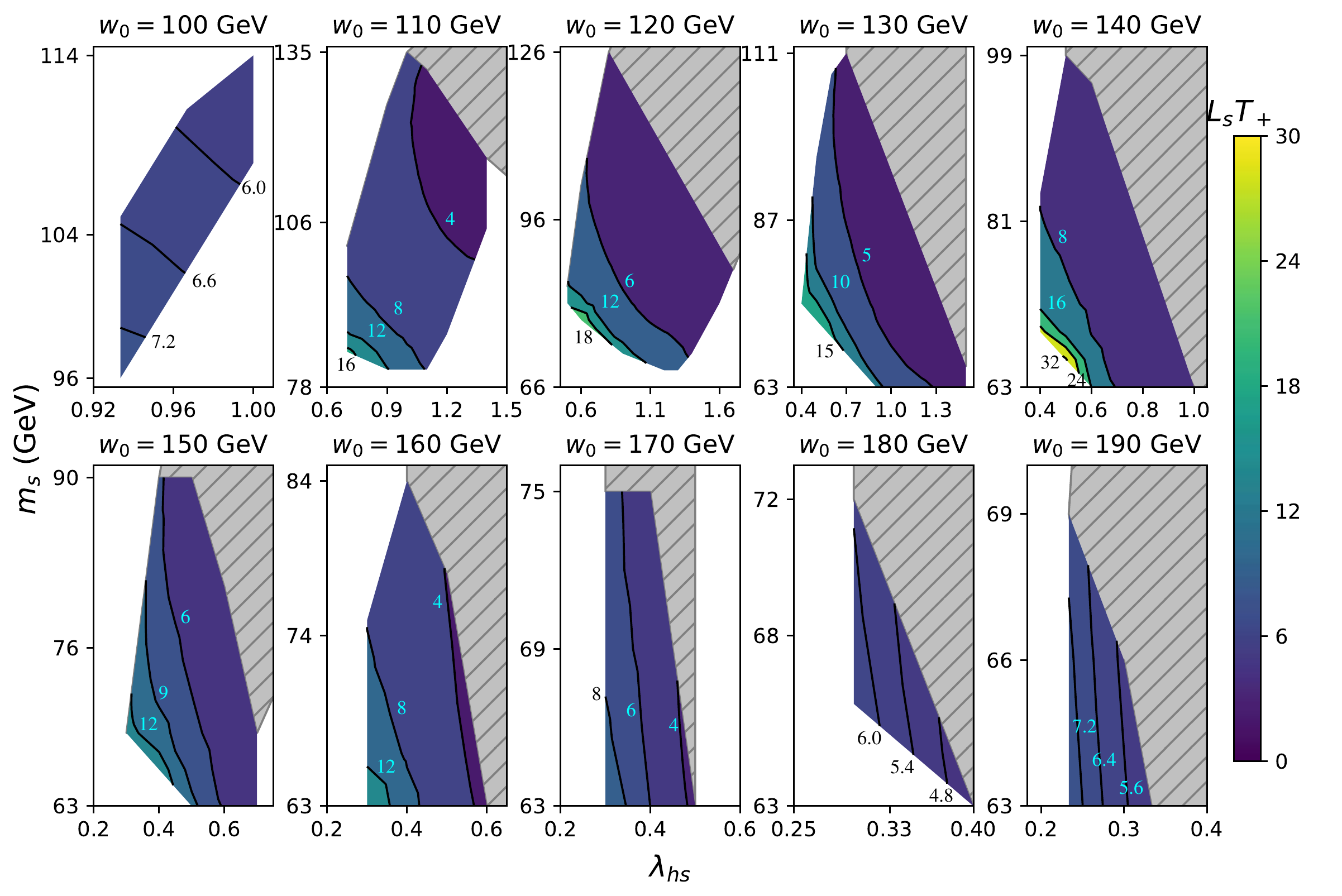}
	\caption{Top two rows (a): Like fig.\ \ref{fig:scan}, but showing the contours of the Higgs wall width $L_hT_+$ where $T_+$ is the temperature in front of the bubble wall.
	Bottom two rows (b): Like fig.\ \ref{fig:scan}, but showing the contours of the singlet wall width $L_s T_+$.
	}
	\label{fig:Lsscan}
	\vskip-17cm 
\leftline{(a)}
\vskip10cm
\leftline{(b)}
\vskip7cm
\end{figure*}

\begin{figure*}[htb]
	\centering
	\includegraphics[width=\textwidth]{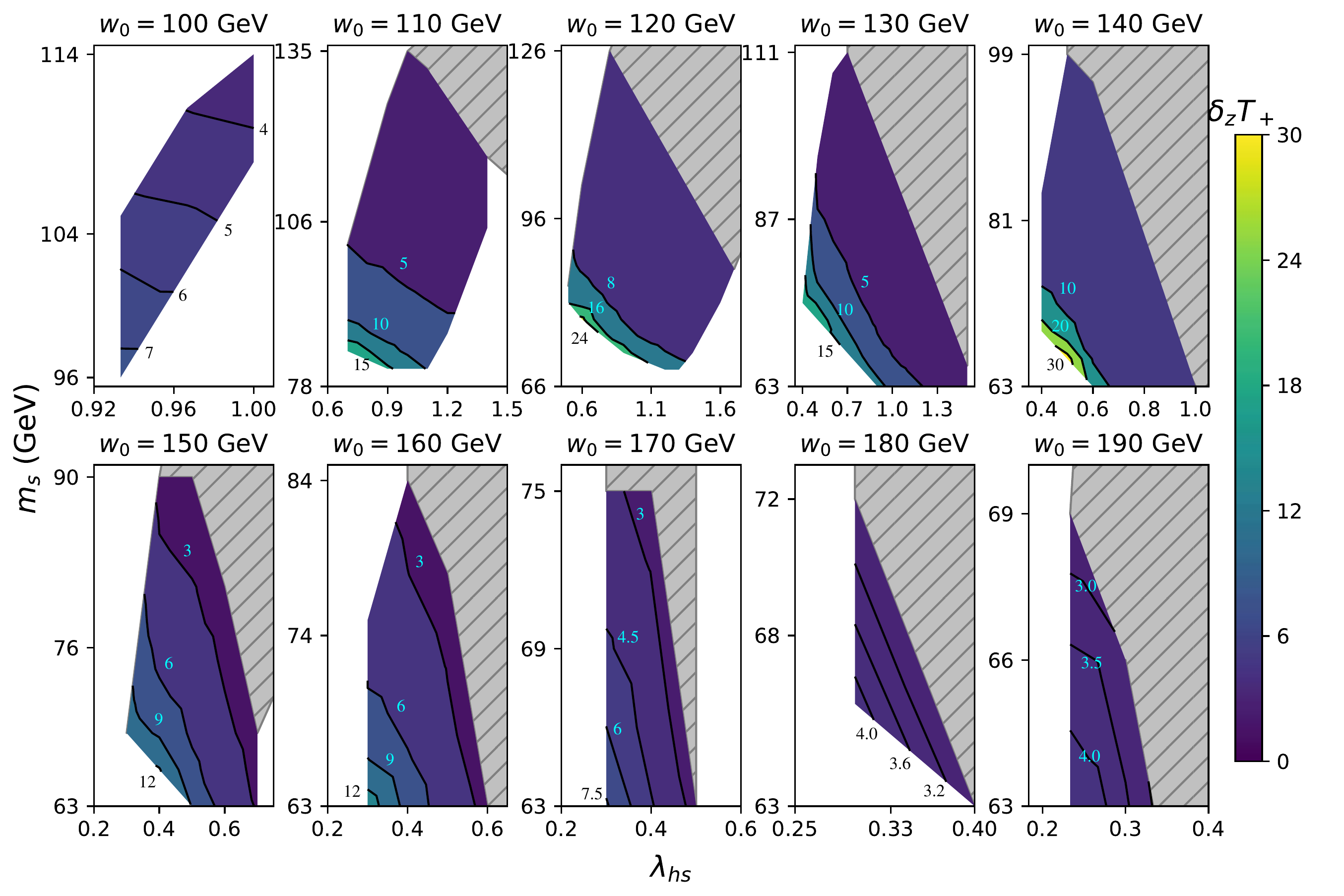}
	\caption{Like fig.\ \ref{fig:scan}, but showing the contours of the wall separation parameter $\delta_z$, defined in eq.\ (\ref{dzeq}).}
	\label{fig:dzscan}
\end{figure*}

\section{Results and Discussion}\label{sec:results}

A scan of the parameter space of the scalar singlet model was performed in the ranges
\bea
\label{scanregion}
0.1 &\le& \lambda_{hs} \le 1.8\nn\\
63\, \textrm{GeV} &\le& m_s \le 138\, \textrm{GeV},\\
100\, \textrm{GeV} &\le& w_0 \le 190\, \textrm{GeV}.\nn
\eea
We did not find viable examples for $w_0 \lesssim 90\,$GeV and only a single point in parameter space that produces subsonic walls for $w_0 \gtrsim 200\,$GeV.  Our results indicate that this covers most of the parameter space of interest for subsonic walls.

We imposed the lower bound $m_s > m_h/2$ so that collider constraints from  invisible Higgs decay ($h~\to~ ss$) do not apply \cite{Aaboud:2019rtt}.  This is a mild restriction, since for $m_s < m_h/2$ and not
too close to the upper limit, these constraints imply the invisible branching ratio is $\lesssim 25\%$, hence $\lambda_{hs} \lesssim 0.01$, which is too small to give rise to a strong phase transition.

\subsection{Wall Velocity Results} \label{sec:vwResults}
Our determinations of the wall speed over the
full parameter space are illustrated in in fig.\ \ref{fig:scan}, showing contours of $v_w$ in the plane
of $m_s$ versus $\lambda_{hs}$, for a series of $w_0$ values. The grey hatched regions indicate
parameters for which no transitions with subsonic
walls were found.
One can see that models with heavier singlets and larger $\lambda_{hs}$ couplings tend to produce faster-moving walls. Generally we find a minimum value for $v_w$, which depends on $w_0$ and is smallest for $w_0\sim 120$\,GeV, where the lowest speed $v_w\cong 0.1$ is found.
The parameters specifying a few benchmark models and their resulting phase transition properties are shown in table \ref{tab:sample}.

\begin{table}[b]
\centering
\begin{tabular}{|c|c|c||c|c|c|c|c|c|}
    \hline
	$m_{s}$ & $\lambda_{hs}$ & $w_0$ &  $v_w$ &  $T_n$ & $T_c$& $v_n/T_n$ & $r$\\
	\hline
	63 & 0.9 & 130 & 0.128 & 103.592 & 104.865 & 2.02 & 1.02 \\
	\hline
	81 & 1.0 & 110 & 0.142 & 124.301 & 125.425 & 1.40 & 1.03 \\
	\hline    
	66 & 0.3 & 160 & 0.306 & 130.532 & 132.677 & 1.28 & 1.05 \\
	\hline
    105 & 0.8 & 110 & 0.426 & 130.646 & 134.461 & 1.24 & 1.11 \\
	\hline
	
\end{tabular}
\caption{Benchmark models with successively faster moving walls. Masses and temperatures are in GeV. $r$ is the measure of supercooling defined in eq.\ (\ref{req}).}\label{tab:sample}
\end{table}

Since it is numerically expensive to compute $v_w$ for a given model 
from first principles, it is useful to look for
relations between it and other quantities
characterizing the strength of the phase transition, that are easier to compute.  In fact we observe a strong correlation between $v_w$ and the double ratio 
\begin{equation}
\label{req}
    r \equiv \frac{v_n/ T_n}{v_c/ T_c}
\end{equation}
where $v/T$ is evaluated
respectively at the nucleation and the critical temperatures.  This is a measure of the degree of supercooling, and its  correlation with $v_w$ is plotted in fig.\
 \ref{fig:supercooling}, showing that $v_w$ increases rapidly with $r-1$.  We find an analytic fit $v_w \cong 0.55\, ( 1 - r^{-13})$, with deviations of order $\pm 0.03$. The maximum value of $r$ found for subsonic bubble walls was $r \cong 1.23$.
It remains close to unity even for
strong transitions, validating the assumption made in section \ref{sec:Temp} that  the equations of state at the nucleation and critical temperatures do not differ significantly from each other.

The fact that a cutoff on $r$ exists, above which it is unlikely to produce subsonic walls, can be seen in fig.\ \ref{fig:rvsTn}, which shows all the models tested, including those found not to have slow bubble walls. It clearly shows that for $r \gtrsim 1.25$, no transitions produce subsonic walls, whereas for $r \lesssim 1.15$ that all the models tested were found to do so.

The impact of the corrections made to the fluids' collision term discussed in section \ref{sec:CollisionTerm} is shown in fig.\ \ref{fig:intRateComparison}. Using the collision term reported in \cite{Moore:1995si} overestimates the wall velocities significantly. For the fastest walls that are subsonic in both calculations, the error is $\sim 20\%$, however, for the slowest walls using the incorrect collision term causes predictions of $v_w$ that is over double the true value. Additionally, since the corrected collision term predicts slower walls, a larger region of parameter space is found to produce subsonic walls than one would predict if using the incorrect collision term. This sensitivity to the collision term implies that further improvements to determining the fluid interaction rates is likely an important future step in improving the accuracy of the $v_w$ calculation.

Fig.\ \ref{fig:scan} shows that subsonic walls require the singlet to be relatively light, $m_s \lesssim
135$\, GeV, often with a relatively large coupling to the Higgs, $\lambda_{hs} \sim 1$.  If $s$
is long-lived enough to escape detection within a collider, Refs.\ \cite{Craig:2014lda,Ruhdorfer:2019utl,Ramsey-Musolf:2019lsf} suggest that
a singlet with these properties may be a target at the high-luminosity LHC, or perhaps more realistically, at a future collider  (from vector-boson fusion production of an off-shell Higgs). On
the other hand, if we take the model at face value, as a complete model with a standard thermal
history, Ref.\ \cite{Craig:2014lda} also finds that the LUX direct
detection experiment \cite{Akerib:2013tjd} rules out
$m_s \lesssim 120$ GeV even though $s$ would make a subdominant contribution to the dark
matter. Of course, additional model ingredients can easily make $s$ unstable on cosmological
time scales without affecting our phase-transition and wall-velocity results. 

\subsection{Wall Shape Results}

Although Fig.\ \ref{fig:wallShape} shows that the wall shapes deviate from a tanh profile, it is nevertheless a useful approximation for concisely encoding
information about the wall shapes.  We have accordingly analyzed our results from the fully numerical algorithm to find the best-fit tanh
profiles, including a possible offset $\delta_z$ between the Higgs and the singlet profiles:
\bea \label{eq:hfit}
    h_\textrm{fit} &=& \frac{h_0}{2}\bigg[1 + \tanh\bigg(\frac{z}{L_h}\bigg)\bigg]\\\label{eq:sfit}
    s_\textrm{fit} &=& \frac{s_0}{2}\bigg[1 - \tanh\bigg(\frac{z - \delta_z}{L_s}\bigg)\bigg] ,
    \label{dzeq}
\eea
where we have allowed for independent widths $L_h$ and $L_s$ of the Higgs and singlet profiles. 

To display results for the wall thicknesses, we have opted to use dimensionless combinations like $L_h T_+$, where $T_+$ is the temperature of the wall.  If one wants to translate these into absolute thicknesses, it can be done using the strong correlation between $L_h T_+$ and $L_h$
in GeV$^{-1}$ units, shown in Fig. \ref{fig:LvsLT}.
Since all models with subsonic walls have nucleation temperatures in the range $70\,\textrm{GeV} \le T_n \le 140\, \textrm{GeV}$
(see Fig.\ \ref{fig:rvsTn}),
and for slow walls the wall temperature does not deviate much from the nucleation temperature, the relationship between these two ways of characterizing $L_h$ is linear with relatively little scatter: $L_h T_+ \cong L_h\; 119$ GeV.  This reflects the fact that the deviations of wall temperature
from the mean value 
$T_+ = 119\,$GeV are relatively small.

\begin{figure}[t]
	\centering
	\includegraphics[width=0.48\textwidth]{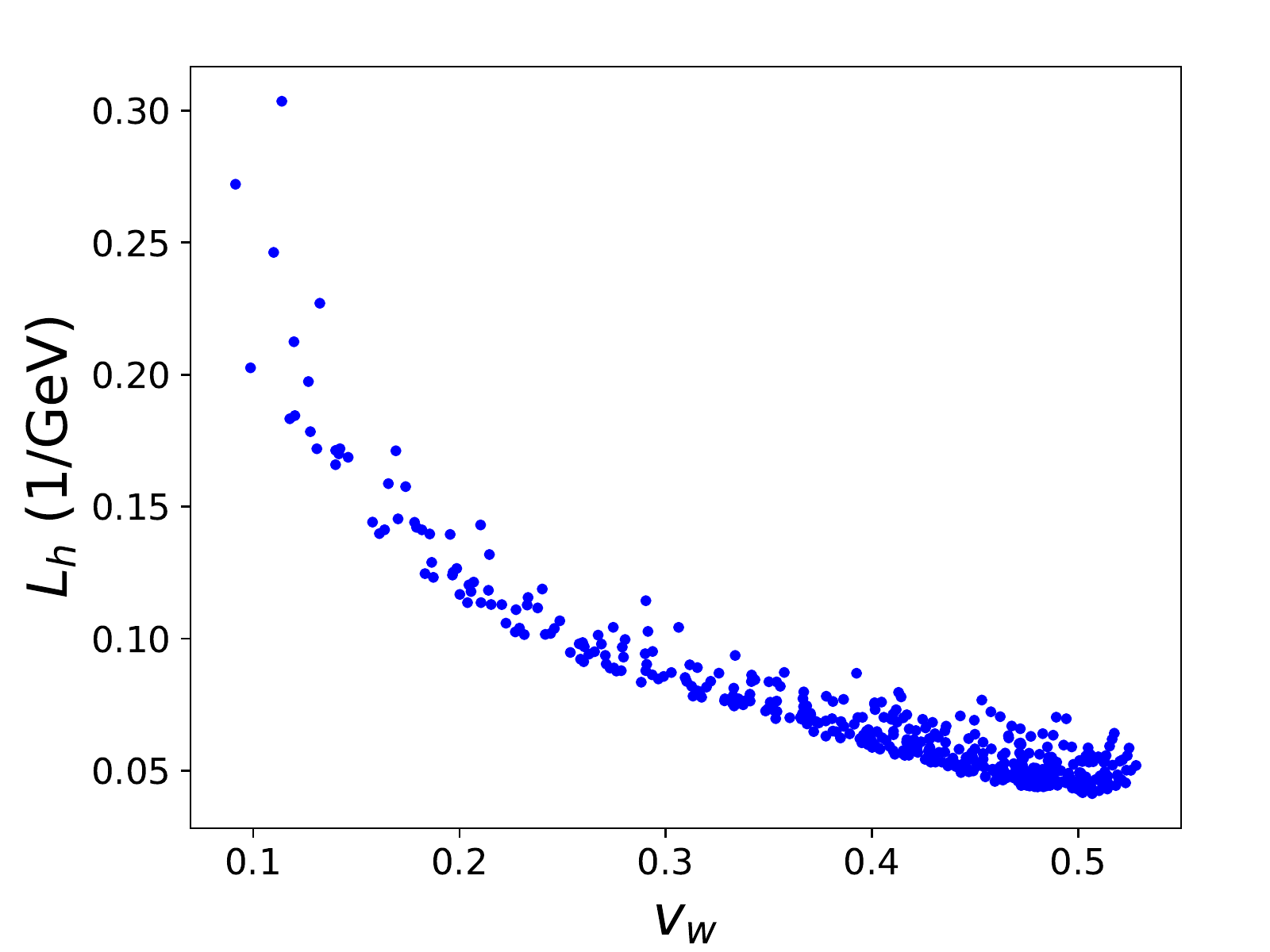}
	\caption{The dependence of Higgs wall width, $L_h$, on the wall velocity, $v_w$.}
	\label{fig:Lhvw}
\end{figure}

\begin{figure}[b]
	\centering
	\includegraphics[width=0.48\textwidth]{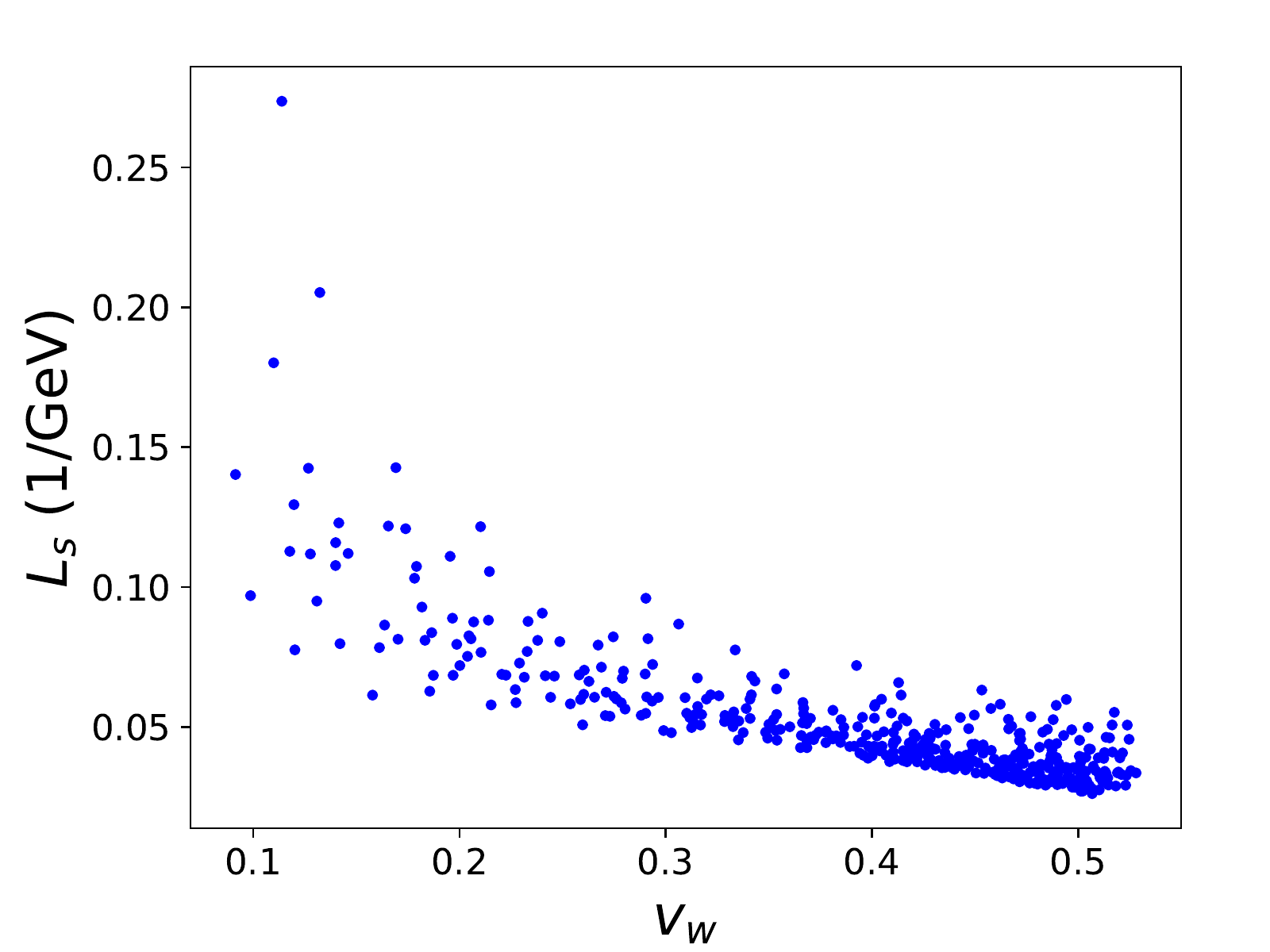}
	\caption{Like fig.\ \ref{fig:Lhvw} but for $L_s$.}
	\label{fig:Lsvw}
\end{figure}

\begin{figure}[htb!]
	\centering
	\includegraphics[width=0.48\textwidth]{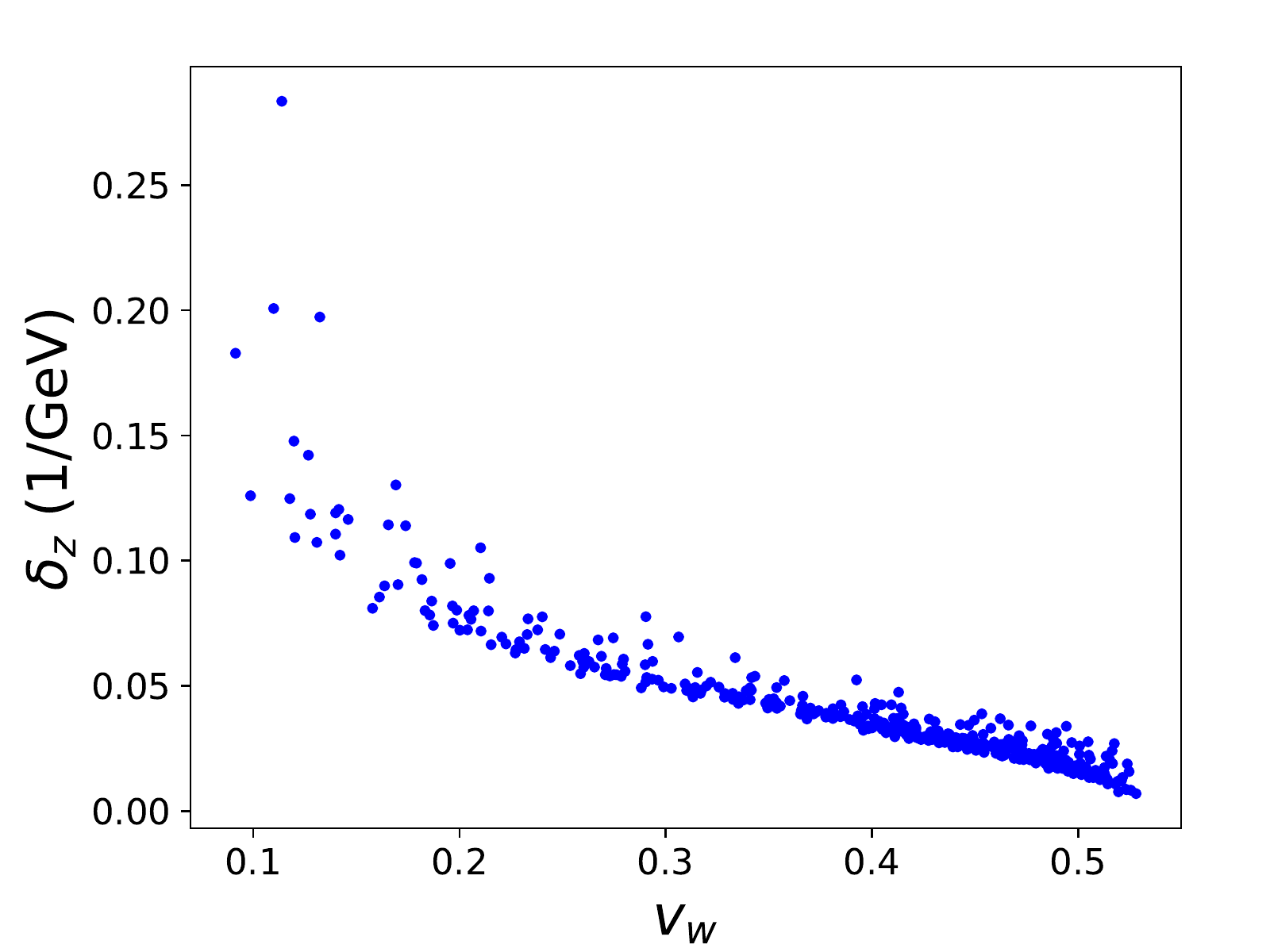}
	\caption{Like fig.\ \ref{fig:Lhvw} but for $\delta_z$.}
	\label{fig:dzvw}
\end{figure}

Contour plots of $L_hT_+$, $L_sT_+$, and $\delta_zT_+$ similar to those for $v_w$ are presented in 
Figs.\ \ref{fig:Lsscan} and \ref{fig:dzscan}.  We find that 
faster walls tend to be thinner and have smaller offsets.  These relationships are plotted in Figs.\
 \ref{fig:Lhvw}-\ref{fig:dzvw}, which show strong correlations, especially in the case of $L_h$. With
rare exceptions, $L_s < L_h$, with $L_s$ typically smaller than $L_h$ by 20-30\%.

\section{Conclusion}\label{sec:conclusion}
This work has laid out a more quantitative methodology than has been previously used, for calculating the wall velocity of bubbles during the electroweak phase transition with an additional scalar field. We improved on previous similar studies by solving for the actual profiles of the scalar fields, rather than just parametrizing them using a $\tanh$ ansatz. Other improvements made here include use of the one-loop Coleman-Weinberg contributions to the potential including the effect of thermal masses, accounting for the sphericity of the bubbles, accounting for the $m/T$-dependence of the $A$-matrix coefficients of the fluid equations, and performing a scan over the three-dimensional parameter space.

Scanning over the parameter space reveals that the scalar singlet model is able to produce slow bubble walls that are preferable for electroweak baryogensis to occur, down to a minimum wall velocity of $v_w \cong 0.1$.  These examples of slow-moving walls only occur in phase transitions with small amounts of supercooling.

There are several ways in which this study can be extended by future work. The precision of the wall velocity calculation could be improved by including additional sources of friction such as from the scalar fields and IR gauge boson modes.   Moreover we have found that the results are rather sensitive to the collision
rates $\Gamma_i$ that enter into the Boltzmann equations for the fluid perturbations.  We have used leading-log
results, that suffer from O(1) uncertainties.  While this work was in progress, Ref.\ \cite{Wang:2020zlf} presented
new results for collision rates beyond the leading-log
approximation, which in some cases exhibited surprisingly large differences from the leading-log counterparts.\footnote{The authors of ref.\ \cite{Wang:2020zlf} are currently trying to better understand and confirm these discrepancies (private communication).}

For a complete analysis of electroweak baryogenesis in the $Z_2$ scalar singlet model, this analysis could be embedded in a more complete model that includes a new source of CP-violation in order to determine the size of the matter anti-matter asymmetry that would be produced. Lastly, extending this analysis to apply for faster walls or even supersonic walls could be of interest for studying other effects of the phase transition such as gravitational waves.   We are currently studying these issues \cite{WIP}.

\section*{Acknowledgments}
We thank B.\ Laurent for useful discussions and comments on the draft, and for providing his updated collision terms.
JC and DTS thank the Aspen Center for Physics for providing a stimulating environment where this work was initiated.
The computations in this work were run using equipment funded by the Canada Foundation for Innovation and supported by the Centre for Advanced Computing at Queen’s University. Besides the previously mentioned CosmoTransitions, the code used for the calculations utilised Eigen \cite{eigenweb} and the GNU Scientific Library \cite{GSLmanual}.  AF and JC are supported by NSERC (Natural Sciences and Engineering Research Council, Canada).

\begin{appendix}
\section{Effective potential}
\label{appA}
The one-loop contribution to the potential can be approximated as 
\begin{equation} \label{eq:CW}
V_1=\!\!\!\!\!\!\sum_{i={h,s,\chi,t,W,Z,\gamma}}\frac{n_i m_i^4(h,s,T)}{64 \pi^2}\bigg[\textrm{ln}\bigg(\frac{m^2_i(h,s,T)}{v_0^2}\bigg) - c_i\bigg]
\end{equation}
where $n_i$ is the number of degrees of freedom of each particle. For the scalar fields, longitudinal $W/Z$ and top quark $c_i = 3/2$ but for the transverse gauge bosons $c_i = 1/2$, in the $\overline{\hbox{MS}}$ scheme. The top quark is the only fermion included in the sum since the contributions from lighter fermions are suppressed by their small Yukawa couplings.
$\chi$ stands for the Goldstone boson contributions.

The one-loop contribution acquires a temperature dependence through the thermal masses of the particles, in this method of carrying out the ring resummation \cite{Parwani:1991gq}. It has been shown that for sufficiently strong phase transitions, a more careful treatment of thermal masses can be important \cite{Curtin:2016urg}.

The scalar masses in eq.\ (\ref{eq:CW}) are given by the eigenvalues of the mass matrix:
\begin{equation}
    M^2_{\textrm{scalar},ij}\equiv \frac{\partial^2V}{\partial\phi_i\partial\phi_j} + m^2_{T,i}\delta_{ij}
\end{equation}
where $\phi_i$ and $\phi_j$ are the five scalar fields summed over in eq.\ (\ref{eq:CW})and
\bea
    m^2_{T,h}  &=& T^2\left(\frac{3g^2 + g'^2}{16} + \frac{y_t^2}{4} + \frac{\lambda_h}{2} + \frac{\lambda_{hs}}{24} \right) \\
    m^2_{T,\chi} &=& m^2_{T,h}   \\
    m^2_{T,s} &=& T^2\left(\frac{\lambda_{hs}}{6} + \frac{\lambda_s}{4}\right).
\eea

The three mass eigenvalues associated with the Goldstone bosons vanish in the vacuum state making those terms in eq.\ (\ref{eq:CW}) formally divergent. This is properly dealt with by introducing a scale coinciding with the Higgs mass, $m_h$, to cut off the IR divergence \cite{Cline:1996mga}.
 
The masses associated with the longitudinal modes of the gauge bosons in eq.\ (\ref{eq:CW}) are given by the eigenvalues of the mass matrix: 
\bea
    M^2_{\textrm{long},ij}&\equiv&
    \begin{bmatrix}
    \frac{g^2h^2}{4} & 0 & 0 & 0 \\
    0 & \frac{g^2h^2}{4} & 0 & 0\\
    0 & 0 & \frac{g^2h^2}{4} & \frac{gg'h^2}{4}\\
    0 & 0 & \frac{gg'h^2}{4} & \frac{g'^2h^2}{4}\\
    \end{bmatrix} \nn\\
    &+& \frac{11}{6}T^2\,\textrm{diag}(g^2,g^2,g^2,g'^2)
\eea
The rest of the field-dependent masses in eq.\ (\ref{eq:CW}) are given by:
\bea
    m_{\textrm{trans},w}^2 &=& \frac{g^2h^2}{4}\nn\\
    m_{\textrm{trans},z}^2 &=& \frac{(g^2+g'^2)h^2}{4}\nn\\
    m_{\textrm{trans},\gamma}^2 &=& 0 \nn\\
    m_t^2 &=& \frac{y_t^2 h^2}{2}
\eea

The counterterm contribution to the potential can be parameterized as
\bea\label{eq:VCT}
V_{CT} &=& \frac{1}{2}\delta m_h^2 h^2 +\frac{1}{2}\delta m_s^2 s^2 + \frac{1}{4}\delta\lambda_h h^4 + \frac{1}{4}\delta\lambda_s s^4\nn\\
&+& \frac{1}{4}\delta\lambda_{hs}h^2s^2 
\eea
The five counterterms were chosen to ensure that the full effective potential at $T=0$ maintains its tree-level values for the scalar masses, potential minima, and scalar mixing. This is done by imposing the following conditions at $T=0$:
\begin{equation}
    \frac{\partial V}{\partial h}\bigg|_{h=v_0,s=0} = \ \ \     \frac{\partial V}{\partial s}\bigg|_{h=0,s=w_0} =0
\end{equation}
\begin{equation}
    \frac{\partial^2 V}{\partial h^2}\bigg|_{h=v_0,s=0} = m_h^2,\quad
     \frac{\partial^2 V}{\partial s^2}\bigg|_{h=v_0,s=0} = m_s^2
\end{equation}
and
\begin{equation}
    \frac{\partial^4 V}{\partial h^2\partial s^2}\bigg|_{h=v_0,s=0} = \lambda_{hs}
\end{equation}
where $m_s = \sqrt{\frac{1}{2}\lambda_{hs}v_0^2 - \lambda_s w_0^2}$ is the mass of the scalar singlet in the true vacuum.

The resulting counterterm parameters are found to be
\begin{equation}
\delta m_h^2 = \bigg(\frac{1}{2} \frac{\partial^2 V_{1}}{\partial h^2} - \frac{3}{2 v_0} \frac{\partial V_{1}}{\partial h}\bigg)\bigg|_{h=v_0,s=0}
\end{equation}
\begin{equation}
\delta m_s^2 = \bigg(- \frac{\partial^2 V_{1}}{\partial s^2} + \frac{v_0^2}{2} \frac{\partial^4 V_{1}}{\partial h^2 \partial s^2}\bigg)\bigg|_{h=v_0,s=0}
\end{equation}
\begin{equation}
\delta\lambda_h = \frac{1}{2 v_0^2}\bigg( \frac{1}{ v_0} \frac{\partial V_{1}}{\partial h} - \frac{\partial^2 V_{1}}{\partial h^2} \bigg)\bigg|_{h=v_0,s=0}
\end{equation}
\begin{equation}
\delta\lambda_s = - \frac{\delta m_s^2}{w_0^2} - \frac{1}{w_0^3}\frac{\partial V_1}{\partial s}\bigg|_{h=0,s=w_0}
\end{equation}
and
\begin{equation}
\delta \lambda_{hs} = - \frac{\partial^4 V_1}{\partial h^2 \partial s^2}\bigg|_{h=v_0,s=0}
\end{equation}

Lastly, the temperature dependence of the potential is given by 
\bea
    V_T &=& - \frac{12 T^4}{2\pi^2}J_F\bigg(  \frac{m_t(h)}{T^2} \bigg)\nn\\
    &+& \sum_{i={h,s,\chi,W,Z}}\frac{n_i T^4 }{2 \pi^2}J_B\bigg(  \frac{m_i^2(h,s,T)}{T^2}   \bigg) 
\eea
where $J_F$ and $J_B$ are functions which describe fermions and bosons temperature-dependent contribution to the one-loop potential. The functions are calculated from
\begin{equation}
    J_F(y) = \int_{0}^{\infty} x^2 \ln\left(1+e^{-\sqrt{x^2 + y^2}}\right)dx 
\end{equation}
and
\begin{equation}
    J_B(y) = \int_{0}^{\infty} x^2 \ln\left(1-e^{-\sqrt{x^2 + y^2}}\right)dx 
\end{equation}

These equations fully describe the one-loop potential of the scalar fields.

\section{Linearized Boltzmann Equations} \label{sec:AppendBoltz}
The following derivation of the linearized moments to the Boltzmann equation, which are used to to determine the friction of the equation of motion, follows closely to that originally expressed in \cite{Moore:1995si}. The difference between that derivation and the one here is that the full dependence of $m/T$ is included here instead of expanding to lowest order. This allows for stronger phase transitions to be quantitatively studied.

As noted in eqs.\ (\ref{eq:fiDef}-\ref{eq:boltzEq}) the fluids are described by the distribution function
\begin{equation} 
f_i(E, z) = \frac{1}{e^{E + \delta_i(z))/T} \pm 1}
\end{equation}
where the $+/-$ is for fermions/bosons and 
\bea 
\delta_i(z) &=& -\Big[T(\delta\mu_i + \delta\mu_{bg})(z) + E(\delta\tau_i + \delta\tau_{bg})(z)\nn\\ &+& p_z(\delta v_i + \delta v_{bg})(z)\Big]
\eea
The background fluid is in chemical equillibrium so for the rest of the derivation $\delta\mu_{bg} = 0$
Deviations from equilibrium in the fluids are governed by the Boltzmann equation
\begin{equation} \label{eq:boltzEq2}
\frac{df_i}{dt} = -C[f_i(E,z)]
\end{equation}

The left side of eq. (\ref{eq:boltzEq2}) can be expanded as
\begin{equation} \label{eq:dtf}
\frac{df_i}{dt} = f_{0,i}'\left(\frac{dE}{dt} + \frac{d\delta_i}{dt}\right)
\end{equation}
where
\begin{equation}
f_{0,i}' \equiv \partial_E f_i|_{\delta_i=0}
\end{equation}

In the fluid's reference frame
\begin{equation}
\frac{d\delta_i}{dt} = \partial_t \delta_i + \frac{p_z}{E} \partial_z \delta_i - \frac{(m_i^2)'}{2E}\partial_{p_z} \delta_i
\end{equation}

Starting with the last term
\begin{equation}
-\frac{(m_i^2)'}{2E}\partial_{p_z}\delta_i = \frac{(m_i^2)'}{2E}(\delta v_i + \delta v_{bg})
\end{equation}
As will be shown the perturbations are sourced by a term proportional to $\frac{(m_i^2)'}{2E} f_{0,i}'$ so terms like the one above which are proportional to $\frac{(m_i^2)'}{2E} f_{0,i}' \delta_i$ are on the same order as $\delta_i^2$ and therefore are ignored to linear order.

This may raise the concern that if $m_i/T$ is not small and $\delta_i \propto \frac{(m_i^2)'}{2E}$, does the linear approximation break down? The $\tanh$ ansatz can be used to set a rough condition on the relation between $v_n/T_n$ and $LT$ under which taking the linear order is valid. That will be derived at the end of this section.

Next one observes that $\partial_t = v_w \partial_z$ in the fluid's reference frame, so to linear order in the perturbations
\begin{equation}
\frac{d\delta_i}{dt} = \left(v_w + \frac{p_z}{E}\right)\partial_z \delta_i\,.
\end{equation}

Going back to eq.\ (\ref{eq:dtf}), the term independent of $\delta_i$ acts as the source term in the perturbations equations.
\bea
\frac{dE}{dt} &=& \frac{d}{dt}(p^2 + m_i^2)^{1/2} \nn\\
     &=& \frac{1}{2(p^2 + m_i^2)^{1/2}}\frac{dm_i^2}{dt} \\
     &=& v_w \frac{(m_i^2)'}{2E}\nn
\eea

Therefore the Boltzmann equation becomes
\begin{equation}
f_{0,i}'\left(v_w + \frac{p_z}{E}\right)\partial_z \delta_i 
+ C[f_i] = -v_w f_{0,i}' \frac{(m_i^2)'}{2E}
\end{equation}
which when expanding out $\delta_i$ it becomes
\bea \label{eq:BoltzPreMoment}
-f_{0,i}'(v_w &+& \frac{p_z}{E})[T\delta\mu_i' + E(\delta \tau_i' + \delta\tau_{bg}') + p_z(\delta v_i' + \delta v_{bg}')]  \nn\\ &+& C[f_i] = -v_w f_{0,i}' \frac{(m_i^2)'}{2E}
\eea

Three moments are taken to turn this into a system of ordinary differential equations. The three moments are $\int \frac{d^3p}{(2\pi)^3}$, $\int \frac{E}{T}\frac{d^3p}{(2\pi)^3}$, and $\int p_z \frac{d^3p}{(2\pi)^3}$.

When taking the first moment, all terms proportional to $p_z$ integrate to zero leaving

\bea \label{eq:moment1a}
\int\frac{d^3p}{(2\pi)^3}&\bigg(&-f_{0,i}'v_w[T\delta\mu_i' + E(\delta \tau_i' + \delta\tau_{bg}')] \nn\\
&-&f_{0,i}' \frac{p_z^2}{E}(\delta v_i' + \delta v_{bg}')   + C[f_i]\bigg)\nn\\ 
&=& \int\frac{d^3p}{(2\pi)^3])}\bigg(-v_w f_{0,i}' \frac{(m_i^2)'}{2E}\bigg)
\eea
Two sets of variabels are then introduced. 
\begin{equation} \label{eq:cdef}
c_j^i = -\int f_{0,i}' \frac{E^{j-2}}{T^{j+1}}\frac{d^3p}{(2\pi)^3}
\end{equation}
and
\begin{equation}\label{eq:ddef}
d_j^i = -\int f_{0,i}' \frac{p^2E^{j-4}}{T^{j+1}}\frac{d^3p}{(2\pi)^3}
\end{equation}
After noting that $p_z^2 = p^2/3$ and substituting eqs. (\ref{eq:cdef}, \ref{eq:ddef}) into eq. (\ref{eq:moment1a}), one gets
\bea
T^4 v_w c_2^i \delta \mu_i' &+& T^4 v_w c_3^i(\delta\tau_i' + \delta\tau_{bg}') \nn\\
&+&T^4 v_w d_3^i(\delta v_i' + \delta v_{bg}')/3 + \int\frac{d^3p}{(2\pi)^3}C[f_i] \nn\\
&=& \frac{T^2 v_w c_1^i (m_i^2)'}{2}
\eea
or after factoring out the $T^4$
\bea\label{eq:moment1}
v_w c_2^i \delta \mu_i' &+&v_w c_3^i(\delta\tau_i' + \delta\tau_{bg}') \nn\\
&+&v_w d_3^i(\delta v_i' + \delta v_{bg}')/3 + \int\frac{d^3p}{(2\pi)^3}\frac{C[f_i]}{T^4} \nn\\
&=& \frac{v_w c_1^i (m_i^2)'}{2T^2}
\eea

The second moment equation is the exact same except with an extra factor of $E/T$ in each term leading to
\bea \label{eq:moment2}
v_w c_3^i \delta \mu_i' &+&v_w c_4^i(\delta\tau_i' + \delta\tau_{bg}') \nn\\
&+&v_w d_4^i(\delta v_i' + \delta v_{bg}')/3 + \int\frac{d^3p}{(2\pi)^3}\frac{E\;C[f_i]}{T^5} \nn\\
&=& \frac{v_w c_2^i (m_i^2)'}{2T^2}
\eea

For the third moment equation, due to the extra factor of $p_z$, the opposite set of terms in eq. (\ref{eq:BoltzPreMoment}) compared to the first two moments integrates to zero leaving
\bea
\int \frac{d^3p}{(2\pi)^3}&\bigg(&-f_{0,i}'\frac{p_z^2}{E}[T\delta\mu_i' + E(\delta \tau_i' + \delta\tau_{bg}')] \nn\\
&-&f_{0,i}' v_w p_z^2(\delta v_i' + \delta v_{bg}')\bigg) = 0
\eea
which becomes
\bea \label{eq:moment3}
d_3^i \delta \mu_i'/3 &+& d_4^i(\delta\tau' + \delta\tau_{bg}')/3 \nn\\
&+&v_w d_4^i(\delta v_i' + \delta v_{bg}')/3 = 0
\eea

As originally stated in section \ref{sec:dynamics},  eqs. (\ref{eq:moment1}, \ref{eq:moment2}, and \ref{eq:moment3}) form a linear system of ODE's that takes the form
\begin{equation}
 A_i (\vec{q_i} + \vec{q}_{bg})' + \Gamma_i \vec{q_i} = S_i
\end{equation}
with $A_i$, $\Gamma_i$, $S_i$, and $q_i$ all taking the same form as they do in section \ref{sec:dynamics}.

Perturbations, $q_i$, are sourced by a term proportional to $\frac{(m_i^2)'}{2T^2}$ so if $\frac{(m_i^2)'}{2T^2} \sim 1$  treating perturbations to linear order is no longer valid. To determine a rough quantitative condition of when this is true the $\tanh$ ansatz can be used where the Higgs wall shape is
\begin{equation}
h(z) = \frac{v}{2}\bigg(\tanh(\frac{z}{LT}) + 1\bigg)
\end{equation}
This conditions will first break down with the top quark which has a mass given by
\begin{equation}
m_t(z)/T = \frac{y_t h(z)}{\sqrt{2}T}
\end{equation}
Then by taking the derivative
\begin{equation}
\frac{(m_t^2)'}{2T^2} = \frac{(\frac{v}{T})^2 y_t^2 \textrm{sech}^2(\frac{z}{LT})\bigg(\tanh(\frac{z}{LT}) + 1\bigg)}{8LT}
\end{equation}
At its maximum value this is equal to
\begin{equation}
\frac{(m_t^2)'}{2T^2}\bigg|_{max} = \frac{4 (\frac{v}{T})^2 y_t^2}{27 LT}
\end{equation}
By ensuring that $\frac{(m_t^2)'}{2T^2}\big|_{max} < 1$ we get the condition
\begin{equation}
\bigg(\frac{v}{T}\bigg)^2 < 6.9 \;LT
\end{equation}

This condition is easily met by all the wall found to have subsonic walls in this paper therefore indicating that the linear order approximation is valid.
\end{appendix}

\bibliographystyle{utphys}
\bibliography{references.bib}

\end{document}